\documentclass[fleqn,usenatbib]{mnras}

\usepackage[T1]{fontenc}

\DeclareRobustCommand{\VAN}[3]{#2}
\let\VANthebibliography\thebibliography
\def\thebibliography{\DeclareRobustCommand{\VAN}[3]{##3}\VANthebibliography}

\usepackage{graphicx}	% Including figure files
\usepackage{amsmath}	% Advanced maths commands
\usepackage{amssymb}	% Extra maths symbols
\usepackage{comment}
\usepackage{soul}
\title[HCO$^+$ and the Effect of Mixing in SN 1987A]{HCO$^+$ and the Effect of Mixing in SN 1987A}

\author[H. M. Davies et al.]{
H. M. Davies,$^{1}$
M. Matsuura,$^{1}$ \thanks{E-mail:  matsuuram@cardiff.ac.uk (MM), GomezH@cardiff.ac.uk (HLG)}
H. L. Gomez,$^{1}$
A. M. S. Richards,$^{2}$
P. Cigan,$^{3, 4}$ \newauthor
R. Indebetouw,$^{5, 6}$
F. D. Priestley,$^{1}$
R. Wesson,$^{1,7,8}$
M. J. Barlow,$^{7}$
J. Larsson,$^{9}$
and
C. Fransson,$^{10}$ 
\\
$^{1}$ Cardiff Hub for Astrophysical Research and Technology (CHART), School of Physics and Astronomy, Cardiff University, \\
Queen's Buildings, The Parade, Cardiff, CF24 3AA, UK\\
$^{2}$ JCBA, School of Physics and Astronomy, University of Manchester, Manchester City, M13 9PL, UK\\
$^{3}$ Department of Physics and Astronomy, George Mason University, 4400 University Dr Fairfax, VA 22030-4444, USA \\
$^{4}$ Celestial Reference Frame Department, United States Naval Observatory, 3450 Massachusetts Ave NW, Washington, DC 20392, USA \\
$^{5}$ Department of Astronomy, University of Virginia, 530 McCormick Road, Charlottesville, VA 22904, USA \\
$^{6}$ National Radio Astronomy Observatory, 520 Edgemont Road, Charlottesville, 22903, VA, USA \\
$^{7}$ Dept. of Physics \& Astronomy, University College London, Gower Street,
London WC1E 6BT, UK \\
$^8$
Department of Physics, Maynooth University, Maynooth, Co Kildare,
Ireland \\
$^{9}$ Department of Physics, KTH Royal Institute of Technology, The Oskar Klein Centre, AlbaNova, Stockholm, SE-106 91, Sweden \\
$^{10}$ Department of Astronomy, Stockholm University, The Oskar Klein Centre, AlbaNova, Stockholm, SE-106 91, Sweden \\
}

\date{Accepted XXX. Received YYY; in original form ZZZ}

\pubyear{2026}

\begin{document}
\label{firstpage}
\pagerange{\pageref{firstpage}$-$\pageref{lastpage}}
\maketitle

\begin{abstract}
We present high angular resolution observations of the HCO$^+$ emission in the central ejecta of the supernova remnant SN 1987A using the Atacama Large Millimeter Array (ALMA). We use this to infer the degree and type of mixing required within the ejecta in order to form HCO$^+$. The distribution of the $J=3-2$ HCO$^+$ emission is co-spatial with that of the $J=2-1$ CO emission, with an overlap between their brightest peaks. The correlation between the two molecules is strong and suggests that HCO$^+$ could form from reactions involving CO. We obtain additional observations of the $J=4-3$ HCO$^+$ emission to calculate the mass of HCO$^+$. The estimated HCO$^+$ mass is $3\text{--}9 \times 10^{-6}\,M_{\odot}$. The relatively large fractional abundance of HCO$^+$ with respect to CO ($M_{\mathrm{HCO}^+}/M_{\mathrm{CO}} = 3 \times 10^{-6}\text{--}3 \times 10^{-4}$) suggests that a moderate amount of hydrogen was mixed into the carbon- and oxygen-rich nuclear zones of the ejecta prior to and during the supernova explosion, in addition to large-scale macroscopic mixing.
\end{abstract}


%%%%%%%%%%%%%%%%%%%%%%%%%%%%%%%%%%%%%%%%%%%%%%%%%%

%%%%%%%%%%%%%%%%% BODY OF PAPER %%%%%%%%%%%%%%%%%%

%10k nh~10^3cm-3  mol clouds Williams & viti book

\section{Introduction}
The explosion of supernova (SN) 1987A in the Large Magellanic Cloud (hereafter LMC) was detected on 23 February 1987 \citep{IAU_SN_discovery_1987}. At a relatively nearby distance of 51.2\,kpc \citep{distance_1991}, its remnant has remained sufficiently bright to be observed across most regions of the electromagnetic spectrum even after three decades of study (see \citealt{Arnett_1989_rev, McCray_1993, McCray, 2025ConPh..66...39B} for reviews).
Within a few months of the explosion, conditions within the ejecta became favourable for the formation of molecules which can help probe the structure and conditions within the ejecta \citep{2026MNRAS.tmp..301W}. The earliest emission by CO was observed 112 days after the supernova explosion (hereafter d$_{\rm SN}$) (e.g. \citealt{1st_CO_ref}), shortly followed by SiO at d$_{\rm SN}\sim$160 (e.g. \citealt{Aitken_1988_1stSiO}). Additionally, H$_2$ was predicted to begin forming around d$_{\rm SN}\sim$15$-$100 and most of it forming around d$_{\rm SN}\sim$ 400$-$1000 \citep{Culhane_McCray_H2_1995, Utrobin_chugai_2005}, but was not detected until d$_{\rm SN}\sim$6500 \citep{2016_fransson_H2_discovery, 2019_Larsson_H2}. The rapid chemical formation routes of these molecules and the physical conditions which leads to their formation are still intensely researched (e.g. \citealt{Petuchowski_1989, R&W, Lepp_1990, Liu_1992, SiO_model, Culhane_McCray_H2_1995, Utrobin_chugai_2005, sarangi_cherchneff, 2018MNRAS.480.5580S, Ono_matter_mixing_2024}).

At present, observations probing the cool molecular regions of the ejecta with ALMA, the Very Large Telescope (VLT), and the JWST have enabled the spatially resolved mapping of molecular distributions. These molecular morphologies provide key insight into the locations and physical conditions under which molecules can exist in the ejecta. Recent studies show that the spatial distributions of CO, SiO, and H$_2$ quite differ \citep{Abellan, Cigan_ref, 2019_Larsson_H2, Matsuura_2024_H2, 2026MNRAS.tmp..301W}. CO and SiO exhibit clumpy structures concentrated toward the centre of the remnant, with SiO being more compact than CO. In contrast, H$_2$ emission displays a bright blob in the south of the ejecta \citep{2019_Larsson_H2, Matsuura_2024_H2}. 
Recent {\it JWST} observations with improved spatial resolution further reveal that the fainter H$_2$ emission follows a more extended distribution, characterized by a distinctive `keyhole' morphology at its centre \citep{Matsuura_2024_H2}. This structure closely resembles the overall distribution of H$\alpha$ emission \citep{Larsson_2013, Larsson_2016}. Despite these similarities, there are subtle differences in peak brightness: the brightest regions are anti-correlated, with H$_2$ emission peaking in the south and H$\alpha$ emission peaking in the west of the ejecta. 
The H$\alpha$ emission is powered by energy deposition from X-rays and UV photons generated by shock interaction with the equatorial ring \citep{H_alpha_Larsson2011, Larsson_2013, Larsson_2016}, whereas the H$_2$ emission is likely powered by UV radiation \citep{Larsson_model_2023}, potentially originating from the ring and/or from the decay of $^{44}$Ti \citep{2016_fransson_H2_discovery}.

HCO$^+$ is a molecular ion which is commonly found in the dense regions of the interstellar medium (hereafter ISM) such as molecular clouds. It has been detected in shock interactions of supernovae with neighbouring molecular clouds \citep{Snell_2005_IC443, W28, W49B}, with the origin of HCO$^+$ attributed to the molecular cloud itself. The first observation of HCO$^+$ in the ejecta of a supernova remnant was by \citet{Mikako_ref} for SN 1987A, which was later followed by a detection in the Crab Nebula \citep{Wootten_crab}. This came as a surprise as chemical network models of supernova remnants either did not include it (e.g. \citealt{sarangi_cherchneff}) or predicted very small abundances for it X$_{\rm HCO^+}\sim$10$^{-18}$ \citep{R&W}. This raises the question of how HCO$^+$ is able to form in SN 1987A in the absence of influences outside the supernova remnant.

HCO$^+$ requires the interaction of hydrogen, carbon and oxygen species to form, with a typical formation process in the ISM being H$_3^+$+CO$\rightarrow$HCO$^+$+H$_2$ \citep{mol_cloud, Oka_H3_06, panessa_23}. 
This proves difficult to achieve in a supernova remnant if the internal nuclear burning zones of the progenitor star are retained post-explosion. This internal structure is comprised of multiple radially-stratified layers of differing elemental compositions arising from the progenitor's nuclear burning zones. This progenitor structure is made up of a core region and an outer envelope, where the envelope contains predominantly hydrogen and an inner helium layer, while the core is comprised of shells of carbon, neon, oxygen, silicon and then to iron at the centre of the core (e.g \citealt{SN87A_abundances_woosley, Shigeyama_presup_model_mixing_1990}). 
In this structure, the hydrogen is separated from the carbon and oxygen regions by a layer of helium, (e.g \citealt{Woosley_6_week_1988, SN87A_abundances_woosley}) hence, disruption of this progenitor structure is needed to trigger HCO$^+$ formation \citep{Mikako_ref}.

Evidence of large-scale disruption to the progenitor stratification after the explosion, came from observations such as the earlier-than-predicted emergence of X-rays from the ejecta (e.g. \citealt{Leising_Share_90}), the `Bochum event' (e.g. \citealt{Utrobin_1995_bochum}) and the clumpy structure of molecules in the ejecta (e.g. \citealt{Abellan, Cigan_ref, 2026MNRAS.tmp..301W}). All of these observations can be explained if large-scale mixing, or macroscopic mixing has occurred.  3D hydrodynamical modelling of the core-collapse supernovae (hereafter CCSNe) found that this macroscopic mixing is caused by Rayleigh-Taylor instabilities which occur at the boundaries of nuclear burning zones \citep{hammer_ref, 3d_model_ref, Utrobin_2015, Utrobin_2019, Utrobin_2021_binary_merger, Gabler_2021, 2026MNRAS.tmp..301W}. This causes large, mushroom/finger-like structures of metals such as Ni, Si, Fe where some may reach velocities {>1000\,km\,s$^{-1}$, which is the average velocity of the material %\textcolor{magenta}{} 
up to and including He in the ejecta (e.g. \citealt{3d_model_ref})}.  These structures traverse through the outer layers of the ejecta, thus disrupting the radial progenitor structure within a few hours after the explosion. A small amount of the hydrogen in the envelope gets inwardly mixed to the inner layers of the ejecta due to some of the hydrogen having slower relative velocities (down to $\sim$ 500\,km\,s$^{-1}$) (e.g. \citealt{3d_model_ref}) compared with the average velocity of the inner zones. %\textcolor{magenta}{ Additionally, similar values for the velocity of both H$\alpha$ and H$_2$ in SN 1987 have been reported by \citet{Kozma_fransson_1998_H2_vel, Larsson_2016, 2019_Larsson_H2}.} 
A second form of mixing, at the atomic level, called microscopic mixing may also occur at the same time as the large-scale mixing which can increase the chances of HCO$^+$ formation. This form of mixing is problematic as the modelling of CO emission at d$_{\rm SN}\leq$1,000 under-produces the observed CO masses if a microscopic mixing scenario is assumed \citep{Lepp_1990, Liu+Dalgarno}.
While 3D hydrodynamic models can predict the macroscopic mixing driven by Rayleigh–Taylor instabilities, their high computational cost makes it difficult to simulate microscopic mixing, which would require resolving spatial scales down to the atomic level. Nevertheless, Rayleigh-Taylor instabilities can generate shear-driven Kelvin-Helmholtz instabilities, as predicted by 2D models \citep{2006A&A...453..661K}. These instabilities promote clump formation \citep{Pinto_woosley_1988, 2006A&A...453..661K}, which can lead to microscopic mixing.
\citet{Nozawa_2003}, who included microscopic mixing in the synthesis of dust species in the ejecta of Population III supernovae, found that microscopic mixing leads to less diverse species of dust grains forming, compared to an unmixed scenario. Both forms of mixing have the potential to aid the formation of HCO$^+$ in the remnant by placing hydrogen, carbon and oxygen in the same location to facilitate chemical reactions.

To answer the question of how HCO$^+$ can form in the ejecta of SN 1987A, we obtained new sub-mm observations of the $J=3-2$ and $J=4-3$ transitions of HCO$^+$ with the ALMA observatory. Using the spatial distribution of the HCO$^+$ emission we investigate the relationship between HCO$^+$ and CO, a potential key element in the formation of HCO$^+$. We analyse the line emission of HCO$^+$ and compute its mass and temperature in order to compare with \citet{Mikako_ref}. We also compare the observations with predictions from hydrodynamic mixing within the remnant.

%%%%%%%%%%%%%%%%%%%%%%%%%%%%%%%%%%%%%%%%%%%%%%%%%%%%%%%%%%%%%%%%%%%%%%%%%%%%%%%%%%%%%%%%%%%%%%%%%
%%%%%%%%%%%%%%%%%%%%%%%%%%%%%%%%%%%%%%%%%%%%%%%%%%%%%%%%%%%%%%%%%%%%%%%%%%%%%%%%%%%%%%%%%%%%%%%%%%

\section{Observations and Data Reduction}
\subsection{HCO$^+$ Data}
\subsubsection{$J=3-2$ HCO$^+$ Observations} \label{sect:HCO_observation_32}

\begin{comment}
\begin{tabular}{ 
|p{1.2cm}|p{1.0cm}|p{1.6cm}|p{1.0cm}|p{1.2cm}|p{1.4cm}|p{1.2cm}|p{1.8cm}|p{1.6cm}|p{2.0cm} }
\multicolumn{9}{|c|}{} \\
 \hline 
line &C &Baseline [m] &$\nu$ [GHz] &PWV [mm] &Angular Scales [$\arcsec$] &N$_{\rm ant}$ &Obs Start &Obs End &SN days \\
 \hline
$J=3-2$ & C40$-$8, C43$-$8, &21.0$-$8547.6  &267.56 &0.3$-$0.7 &0.03$-$13.39 &38$-$49 &2017-09-05 &2019-07-28 &11,153$-$11,844   \\

$J=3-2$ & C43$-$4 &15.1$-$1261.4  &267.56 &0.9$-$1.5 &0.22$-$18.62 &43$-$45 &2018-09-10 &2018-09-14 &11,522$-$11,526   \\ 
$J=4-3$ &C43$-$3 C43$-$5 &15.1$-$1713.8  &356.73 &0.6$-$0.9 &0.12$-$14.05 &48$-$49 &2018-08-31 &2018-10-20 & 11,513$-$11,562  \\

 %CO      &$J=2-1$  & & 230.54    & 10,054, 10,479 
 %to 10,494 \\
 %SiO     &$J=5-4$  & & 217.11    & 10,054, 10,479 to 10,494 \\
 \hline
\end{tabular}
\end{comment}

\begin{table*}
\begin{tabular}{ 
|p{1.2cm}|p{1.0cm}|p{1.6cm}|p{1.0cm}|p{1.2cm}|p{1.4cm}|p{1.2cm}|p{1.8cm}|p{1.6cm}|p{2.0cm} }
\multicolumn{9}{|c|}{} \\
 \hline 
HCO$^+$ Line &C &Baseline [m] &$\nu$ [GHz] &PWV [mm] &Angular Resolution [$\arcsec$] &N$_{\rm ant}$ &Obs Start &Obs End &SN Days, d$_{\rm SN}$ \\
 \hline
$J=3-2$ & C40$-$8 &21.0$-$3696.9 &267.56 &0.3 &0.08 &38 &2017-09-05 &2017-09-05 &11,153 \\

$J=3-2$& C43$-$8 &92.1$-$8547.6  &267.56 &0.3$-$0.7 &0.03 &47$-$49 &2017-11-21 &2017-11-23 & 11,229$-$11,231   \\

$J=3-2$ & C43$-$4 &15.1$-$1261.4  &267.56 &0.9$-$1.5 &0.22 &43$-$44 &2018-09-10 &2018-09-14 &11,522$-$11,526   \\ 
$J=3-2$& C43$-$8 &92.1$-$8547.6  &267.56 &0.5 &0.03 &44 &2019-07-21 &2019-07-21 & 11,837  \\ 
$J=3-2$& C43$-$8 &92.1$-$8547.6  &267.56 &0.4 &0.03 &45 &2019-07-28 &2019-07-28 & 11,844  \\ 

 \hline
$J=4-3$ &C43$-$5 &15.1$-$1713.8 &356.73 &0.6 & 0.12 &48 &2018-10-20 &2018-10-20 &11,562 \\

$J=4-3$&C43$-$3& 15.1$-$~~783.5 &356.73  &0.9 &0.27 &49 &2018-08-31 &2018-08-31 &11,512 \\  

 %CO      &$J=2-1$  & & 230.54    & 10,054, 10,479 
 %to 10,494 \\
 %SiO     &$J=5-4$  & & 217.11    & 10,054, 10,479 to 10,494 \\
 \hline
\end{tabular}

\caption{The HCO$^+$ observations analysed in this work. C is the configuration name of the ALMA antennas, the configuration of antennas affects the maximum and minimum baseline length between antennas and therefore the angular resolution of the observations. $\nu$ shows the transition frequencies of HCO$^+$ in the rest frame, PWV quotes the range of precipitable water vapour at the time of observation, N$_{\rm ant}$ is the minimum and maximum number of antennae used. The final three columns state the start and end dates of the observations, and time in days since the initial supernovae explosion. %\hd{removed max values of the ang.res, this was actually the (incorrectly calculated) MRS but honestly I don't think we need to show it. Datacubes are roughly 1 arcsec across, and MRS minimum is approx 2.5 arcsec. We're not really interested in capturing the entire scale of 87A just the ejecta, the anglular resolution is more important for this study.}
}
\label{table:observations}
\end{table*}

We obtained spatially resolved line emission from HCO$^+$ $J=3-2$ at 267.56\,GHz using the ALMA Observatory (project number 2016.1.00077.S in the ALMA Archive). %\textst{Line emission was collected using ALMA's band 6 recievers.} 
Interferometric observations spanned from the 5th of September 2017 to the 28th of July 2019 (SN days, ${\rm d_{SN}}$, 11,153 and 11,844, and ALMA observation Cycles 4 to 6) and consisted of using the ALMA antenna array configurations of C40$-$8, C43$-$8 and C43$-$4. The former two configurations are classed as `extended' arrays which have baselines ranging from  21.0--3696.9 and 92.1--8547.6\,m. C43$-$4 is a compact configuration, with baselines ranging from 15.1--1261.4\,m. This equates to an integration time of roughly 5\,hours for the extended array configurations and 2.5\,hours for the compact array configuration. Further details of the observations can be found in Table~\ref{table:observations}. Calibration of the data was performed using the ALMA pipeline. All configurations used quasi-stellar objects for calibration of the bandpass, phase and flux scales. These were: J0635$-$7516 and J0522$-$3627 for the bandpass calibration, J0601$-$7036 and J0529$-$7245 were used to calibrate the phase and J0519$-$4546 was used to calibrate the flux scale.

The $J=3-2$ HCO$^+$ observations spanned over two adjacent `spectral windows', combined they range from 265.6--269.2 GHz, which covers the $J=3-2$ HCO$^+$ transition frequency which, due to the expanding ejecta, its line emission has been Doppler broadened to have full-width at half maximum, hereafter FWHM, of $\sim$2000\,km\,s$^{-1}$ \citep{Mikako_ref}. A non-zero continuum level is also present in these data, which we will identify and subtract from the HCO$^+$ data in Sect.\,\ref{section:LE_measurements}.

The data were imaged using the Common Astronomy Software Applications' ({\sc CASA}, \citealp{CASA}) {\sc tclean} function and weighted using the `Briggs' weighting option with a robust value of 0.5. This ensured a reasonable signal-to-noise ratio while also keeping some small-scale structures present within the ejecta that would otherwise be blurred from the natural weighting option. The current configuration and {\sc tclean}'s weighting resulted in a FWHM beam size of 0.085$\arcsec\times$0.072$\arcsec$ with a beam position angle of 44.312$^{\circ}$. The root mean square, hereafter RMS, noise of the images, was determined via an off-source aperture and systematic uncertainties in flux measurements is assumed to be 7\,$\%$ at these observing bands\footnote{\url{ https://almascience.nrao.edu/documents-and-tools/cycle5/alma-technical-handbook }}.

Due to the long time-frame of observations, spanning $\sim$ 2 years, we check whether the ejecta morphology may have significantly changed in this period. First we check the change in the dynamical expansion of the ejecta. Assuming a velocity of 2000\,km\,s$^{-1}$, the angular expansion predicted between the observation start and end dates would be $\le 0.02\arcsec$. This is smaller than the beam size of the observations. Second, we check the change in potential heating sources within this period. Assuming the dominant heating source in the inner-most ejecta is via $^{44}$Ti decay \citep{Matsuura_2011, Mikako_dust_2015, Jones_2023}, we predict the resulting decrease in thermal energy in the ejecta across the $\sim$ 690 days between the first and last observation is $\sim$2.2\,$\%$. This is assuming the decrease in heating is proportional to the decay rate of $^{44}$Ti where its half-life is 58.9 years \citep{Ahmad_2006_Ti_hlife}. Additionally, the total of X-ray deposition varies by $\sim$3.8\,$\%$ across the observation period \citep{Ravi_X-ray_2024} therefore we conclude that only minimal changes to the ejecta are expected during the time period of our observations.

\subsubsection {$J=4-3$ HCO$^+$ Observations} \label{HCO+4-3obs}
To determine physical properties of the HCO$^+$ emission, measurements of more than a single line are needed. %\textst{an adjacent HCO$^+$ transition is needed}. 
Therefore we also use the HCO$^+$ $J=4-3$ transition at 356.73\,GHz (project code 2017.1.0221.S in the ALMA archive). Emission from the HCO$^+$ $J=4-3$ transition was collected using ALMA's band 7 receivers at an ALMA antenna configuration C43$-$5 and C43$-$3 with baselines spanning from 15.1 to 1713.8\,m and 15.1 to 785.5\,m respectively. The data were collected on the 31st of August 2018 and the 20th of October 2018 (SN days 11,513 and 11,562 respectively) with an integration time of roughly 1.2 hours. The flux was calibrated using J0529$-$7245, systematic uncertainties in the flux is assumed to be 7\,\% for ALMA observing band 7 \citep[ALMA Technical Handbook 5][]{ALMA_Tech_5}. RMS uncertainties were determined in the same manner as the RMS in the $J=3-2$ HCO$^+$ data cube.

%\textcolor{red}{Haley: in what - aperture/pixels/background - need to be specific}
%{\mm {(Whether target is resolved or not should appear later, only after the image has been shown. Not yet mention in observational section. Also unclear what is "zero-level continuum". Rephrased) 
The FWHM beam size was 0.340$\arcsec\times0.265\arcsec$ with a position angle of $-10.7^{\circ}$. This observation also consists of two `spectral windows' where their combined frequency range spans 354.8 to 358.00\,GHz ($-$1350 to 1650\,km\,s$^{-1}$ with respect to the rest frequency of the $J=4-3$ HCO$^+$ at 356.73\,GHz), which cover only the broadened HCO$^+$ $J=4-3$ emission centred around its transition frequency whereas we expected a similar velocity range to that of the $J=3-2$ HCO$^+$ transition of $\sim$±2000\,km\,s$^{-1}$ \citep{Mikako_ref}. A non-zero continuum-level is also present in the $J=4-3$ HCO$^+$ emission and this level is estimated in Section\,\ref{section:LE_measurements}.

\subsection{Ancillary data}

\subsubsection{CO and SiO Observations}
Observations for the $J=2-1$ CO line at 230.54\,GHz and the SiO $J=5-4$ line at 217.11\,GHz were collected on SN days 10054 (September 2014) and 10479--10494 (November 2015), and reported by \citet{Abellan}, \citet{Cigan_ref} and \citet{2026MNRAS.tmp..301W}. %These molecular emissions were observed in ALMA Cycles 2 and 3 in September 2014 and November 2015 respectively. 
%Further information regarding the CO and SiO data can be found at \citet{Abellan}. 
We use the velocity channel maps of CO and SiO from \citet{Cigan_ref}. Their original FWHM beam sizes and position angles were 0.06$\arcsec\times$0.04$\arcsec$ and 27.43$^\circ$ and 19.74$^\circ$ for CO and SiO respectively \citep{Cigan_ref}. 
%To ensure comparison is fairer between HCO$^+$, CO and SiO observations, 
We re-convolved these images to the coarser beam size of 0.085\,$\arcsec\,\times\,$0.072\,$\arcsec$, identical to that of the $J=3-2$ HCO$^+$ channel maps, using %This smoothing process was done through the
{\sc CASA} function {\sc imsmooth}. 
The change in dynamical expansion between these observations and the HCO$^+$ observations, following the same calculations for it in Sect.~\ref{sect:HCO_observation_32}, results in a maximum predicted angular expansion of $\sim$0.04$\arcsec$, which is less than half of the beamsize, hence, the impact is not significant.

\subsubsection{H$\alpha$ Observations}
The {\it Hubble Space Telescope} observed SN\,1987A with the Wide Field Camera 3 (WFC3) F625W filter; the emission in this band is dominated by H$\alpha$ \citep{H_alpha_Fransson2015, Larsson_2016}. We used the image from \citet{2019ApJ...886..147L} which was acquired on 2018 July 7th (SN day: 11,458) with an exposure time of 1200\,s. The angular scale of each pixel in the image is $\approx$ 0.025$\arcsec$. %The HST F625W band image was aligned to the ring as observed by ALMA at 315\,GHz (described in detail in the appendix of \citealt{Cigan_ref}). 
% \mm{(Where is this resolution from? The HST pixel scale is 0.025”, which is under-sampled, and it is usually sampled multiple times to increase the resolution better than that value.)

\subsubsection{H$_2$ Observations}
{\it JWST} observed SN~1987A using the NIRCam F212N filter, which predominantly traces the $v=1$--$0$ S(1) H$_2$ emission in the ejecta. This filter also captures primarily \ion{He}{I} emission from the equatorial ring. The observations were carried out on 2022 September 1 and 2, corresponding to days 12\,974 and 12\,975 after the supernova explosion was first detected. Further details of the observations can be found in \citet{Matsuura_2024_H2}.

\subsubsection{315\,GHz Continuum Observations}
ALMA observed the continuum at 315\,GHz on 2021 November 2nd (supernova day 12671) \citep{Matsuura_2024_H2}. 
The frequency coverage was 313.73--317.52 GHz (955.57--944.17~$\mu$m) in kinematic local standard of rest.
The FWHM beam size and position angle of the image is  0.081$\arcsec\times$0.068$\arcsec$ and 33.66$^\circ$, , providing the highest angular resolution obtained for dust continuum observations with ALMA. To aid comparison to the $J=3-2$ HCO$^+$ images, this beam size was re-convolved to the coarser HCO$^+$ FWHM beam of 0.085\,$\arcsec\,\times\,$0.072\,$\arcsec$. %For further information, see \citet{Matsuura_2024_H2}. 

The total continuum flux density of the ejecta (i.e. excluding the ring) is $1.7 \pm 0.2$~mJy. The measurement was made within an elliptical aperture of $0.51\arcsec \times 0.58\arcsec$.
The uncertainty was estimated from the rms flux, using apertures of the same size placed outside the equatorial ring, representing only the statistical error and excludes systematic contributions (e.g. flux calibration and aperture definition). This flux density is consistent with the value measured on 2015 July 25 at 315.68~GHz \citep{Cigan_ref}.

%%%%%%%%%%%%%%%%%%%%%%%%%%%%%%%%%%%%%%%%%%%%%%%%%%%%%%%%%%%%%%%%%%%%%%%%%%%%%%%%%%%%%%%%%%%%%%%%%%%%%%%%%%%%%%%%%%%%%%%%%%%%%%%%%%%%%%%%%%%%%%%%%%%%%%%%%%%%%%%%%%%%%%%%%%%%%%%%%%%%%%%%%%%%%%%%%%%%

\section{Image Analysis}
\subsection{Estimating the Contribution of the Continuum in the HCO$^+$ Images} \label{sect:Continuum_estimation}

%Dust continuum comparison
\begin{figure}
    \includegraphics[clip,  width=\columnwidth]{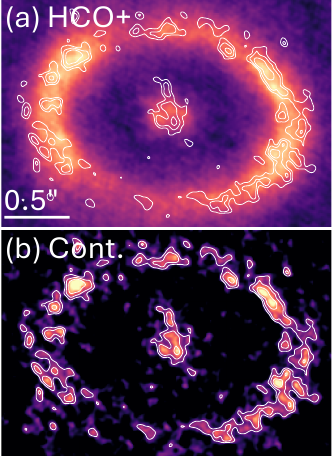}
    \caption{(a) An image of the brightness distribution of the continuum-included $J=3-2$ HCO$^+$ emission, integrated across its velocity space. Overlain are contours of the 315\,GHz continuum emission from \citet{Matsuura_2024_H2}, which is displayed in panel (b). %which has been smoothed to the coarser HCO$^+$ beam size. 
     The continuum consists of dust emission from the ejecta and synchrotron emission from the equatorial ring.
     The continuum contour levels correspond to 4.5$\times$10$^{-5}$, 6$\times$10$^{-5}$\,Jy\,beam and 8$\times$10$^{-5}$\,Jy\,beam, where the beam size is the original 0.081\arcsec$\times$0.068\arcsec. % with the position angle of 34$^{\circ}$. %, where the former (lower) level corresponds to the mean threshold (3$\times$RMS) used in the pixel-by-pixel correlations analysis in Sect.~\ref{Sect:HCO_and_CO_comp}. 
      The dust emission peaks where there is lower HCO$^+$ brightness. }
    \label{fig:dust_comparison}
\end{figure}

%Dust SED figure
\begin{figure}
	\includegraphics[clip, trim=0.2cm 0.1cm 0.3cm 1.1cm, width=\columnwidth]{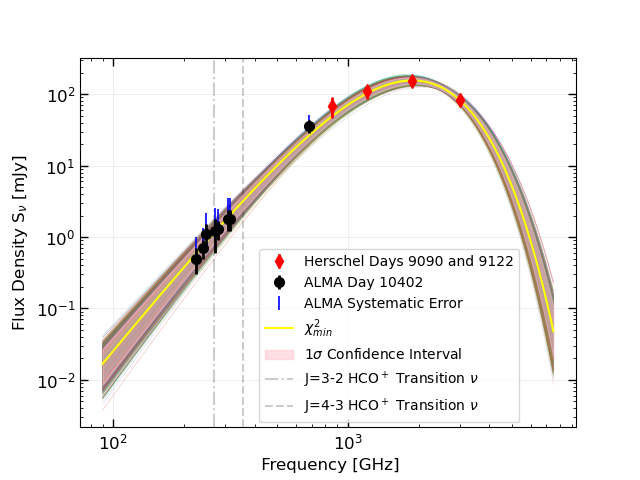}
    \caption{A modified blackbody fit to sub-millimeter and far-infrared observations of thermal dust emission using a Monte-Carlo simulation. Circles show ALMA dust observations in the sub-mm where the temporal span of observations is averaged to SN day 10402 \citep{Cigan_ref}. Diamonds show {\it Herschel} dust observations in the far-infrared taken on SN days 9090 and 9122 \citep{Mikako_dust_2015}. The blue error bar on the ALMA data indicates an additional systematic error which was not included as uncertainty in the MC simulation and fit. The yellow line indicates the line of best fit for the continuum level. The pink region indicates the confidence interval which captures 1$\sigma$ of the distribution of all three parameters which were varied in the MC. The grey dashed lines indicate the HCO$^+$ transition frequencies corrected for the reference frame of SN 1987A at $v_{\rm helio}=$287\,km\,s$^{-1}$ \citep{Sys_vel_1987A_2008}.}
    \label{fig:therm_dust}
\end{figure}

As mentioned before, there is a non-zero continuum level present at the frequencies of our HCO$^+$ transitions in the ejecta. We investigate what effect this continuum level has on the distribution of our HCO$^+$ emission and then we estimate and subtract the continuum levels from the HCO$^+$ line emission. The expected breadth of the HCO$^+$ line (at FWHM of 2000 \,km\,s$^{-1}$) \citep{Mikako_ref} meant there were no adjacent, line-free channels to use for estimating and subtracting the continuum levels from our HCO$^+$ data cubes. Assuming the continuum levels are predominantly emission from dust \citep{Cigan_ref}, we instead compare the spatial distribution of dust emission in the 315\,GHz image to that of the $J=3-2$ HCO$^+$ emission which is seen in Fig.~\ref{fig:dust_comparison}. Note that we integrate the HCO$^+$ velocity channel map across the velocity range $-$1050 to 1650\,km\,s$^{-1}$ to give the distribution in the figure. Superimposed on the HCO$^+$ image are the contour levels of the dust emission at 7$\times$10$^{-5}$ and 9$\times$10$^{-4}$\,Jy\,beam$^{-1}$. The contour levels corresponds to roughly the lower and upper limits of $\sim$3$\times$\,RMS in the channels of the HCO$^+$ emission ($\sim$7.5$-$10.5$\times$10$^{-5}$\,Jy\,beam$^{-1}$). The peak of the 315\,GHz dust emission sits to the south, where there is lower HCO$^+$ emission. We conclude that the spatial distribution of the continuum has little impact on the distribution of the continuum-included HCO$^+$ as the peaks of the brightest intensities of the dust and HCO$^+$ do not overlap. Therefore, our comparisons of the spatial distribution CO, SiO and H$\alpha$ with that of HCO$^+$ are valid, despite there still being continuum included in the $J=3-2$ HCO$^+$ emission.

The continuum levels present in the HCO$^+$ line emission which is estimated later on, will overestimate the HCO$^+$ emission and therefore the column density and mass estimates of HCO$^+$, therefore the continuum levels need to be estimated and subtracted from the HCO$^+$ line emission.
%\textcolor{green}{The following has just been re-shuffled from other sections.}\\
The continuum levels were estimated by fitting a modified blackbody dust curve to thermal dust emission observations from \citet{Mikako_dust_2015} and \citet{Cigan_ref}, shown in Figure~\ref{fig:therm_dust}. These observations comprised of far-IR {\it Herschel} observations ranging from 3000 to 857.4 GHz taken on SN day 9090 and 9122 \citep{Mikako_dust_2015} and sub-mm ALMA observations ranging from 225.5 to 679.2 GHz collected from SN days 10352 to 10441 \citep{Cigan_ref}.
The sub-mm observations are resolved enough to measure the dust emission from the ejecta. Despite the ejecta and ring not being resolved for the dust observations in the far-IR, we assume a negligible contribution of dust emission from the ring at these wavelengths, due to the ring dust fluxes peaking in the mid-IR (\citealt{Mikako_dust_2015, 2024ApJ...965...51B} and references therein). The modified blackbody dust curve is defined as $S_{\nu}=M_d\kappa_{\rm abs}B_\nu(T)/d^2$, where $S_{\nu}$ is the thermal dust emission, $M_d$ is the dust mass in kg, $\kappa_{\rm abs}$ is the mass absorption coefficient in m$^2$kg$^{-1}$, $B_\nu(T)$ is the Planck function at a single temperature $T$, and $d$ is the distance to SN 1987A in m. Since properties of the dust such as grain size and composition are unknown, the dust mass absorption coefficient can be simplified by a power-law, $\kappa_{\rm abs}=\kappa_0(\lambda/\lambda_0)^{-\beta}$. Where $\kappa_0$ is the mass absorption coefficient at a reference wavelength, $\lambda_0$, these were 0.07\,m$^2$kg$^{-1}$ and 850\,$\mu$m, taken from \citet{James_2002}. 
To estimate robust uncertainties from the thermal dust observations, we use a Monte Carlo (hereafter MC) error estimation. We perturb the flux densities 5000 times assuming a normal distribution centred on the observed flux, and a $1\,\sigma$ width equal to the error in the flux. The uncertainties for the sub-mm flux densities in Figure~\ref{fig:therm_dust} are listed in \citet{Cigan_ref}; for this analysis we include uncertainties from the flux calibration, RMS and contamination of flux from the ring in continuum ejecta observations.
 The uncertainties of the far-IR observations arise from calibration errors and the uncertainty in the background subtraction \citep{Mikako_dust_2015}. An additional uncertainty on the positive side of the sub-mm fluxes arising from a discrepancy between Cycle 0 and 2 observations \citep{Cigan_ref}, shown as a blue error-bar in Figure~\ref{fig:therm_dust}, is not included in the MC analysis due to it having little effect on the variability of the MC modified blackbody fits to the data. The MC modified blackbody fits are ultimately limited by the tight constraints provided by the far-IR data. 
 
 Each of the 5000 dust spectral energy distributions, (hereafter SEDs) produced by the MC and their minimum of a goodness-of-fit chi-square, $\chi^2$, values were calculated. Uncertainties in the fit parameters were estimated by adopting a confidence interval of $\chi^2_{\rm min}+\Delta\chi^2$, where $\Delta\chi^2$ = 3.5 \citep{avni_1976_chi2}, which captures the 1$\sigma$ distribution whilst considering our three parameters ($\beta$, $T$ and $M_d$) simultaneously. The result of this confidence interval is seen as the pink-shaded regions in Figures~\ref{fig:therm_dust} and \ref{fig:FD_plots}. The best-fitting parameters from the thermal dust emission and their uncertainties were found to be: $\beta$=2.1$^{+0.3}_{-0.3}$, $T$=17.8$^{+1.6}_{-1.4}$\,K and $M_d$=1.4$^{+0.3}_{-0.3}\,M_\odot$. Based on these results, we estimate that the continuum levels from dust emission at the two HCO$^+$ frequency ranges is (2.69±0.29)$\,\times10^{-20}$\,W\,m$^{-2}$ for $J=3-2$ transition and (1.02±0.92)$\,\times10^{-19}$\,W\,m$^{-2}$for $J=4-3$ transition. 

One caveat in this analysis is the use of the {\it Herschel} far-IR data which was observed roughly 7 years before our HCO$^+$ observations. We investigate whether this affects our continuum levels at the HCO$^+$ transition frequencies by estimating the decrease in the {\it Herschel} Far-IR data across the 7-year time difference using calculations described in \citet{Cigan_ref}. Assuming the dominant heating source of the ejecta dust is the decay of $^{44}$Ti \citep{Matsuura_2011}, which has a half life of 58.9 years \citep{Ahmad_2006_Ti_hlife}, the predicted decrease in the decay energy of $^{44}$Ti between {\it Herschel} and the end of our observations at 11,844 days, is proportional to the decrease in luminosity of the modified blackbody. We can quantify the resulting decrease in temperature by using the Stefan-Boltzmann law and assuming an initial 20\,K temperature results in a temperature decrease to 19.54\,K. This results in a decrease in the continuum level of 3.1$-$3.4$\%$ at the frequency ranges of the HCO$^+$ transitions centred at $\sim$\,267 and $\sim$\,356\,GHz of our modified blackbody thermal dust emission curve. This change in the continuum level is well within our uncertainties of our MC fits to the modified blackbody.

\subsection{Comparison of HCO$^+$, CO and SiO Spatial Distributions} \label{Sect:HCO_and_CO_comp}

%CO contour chan map
\begin{figure*}
	\includegraphics[clip, trim=0.5cm 0.9cm 0.1cm 1.9cm, width=19.5cm]{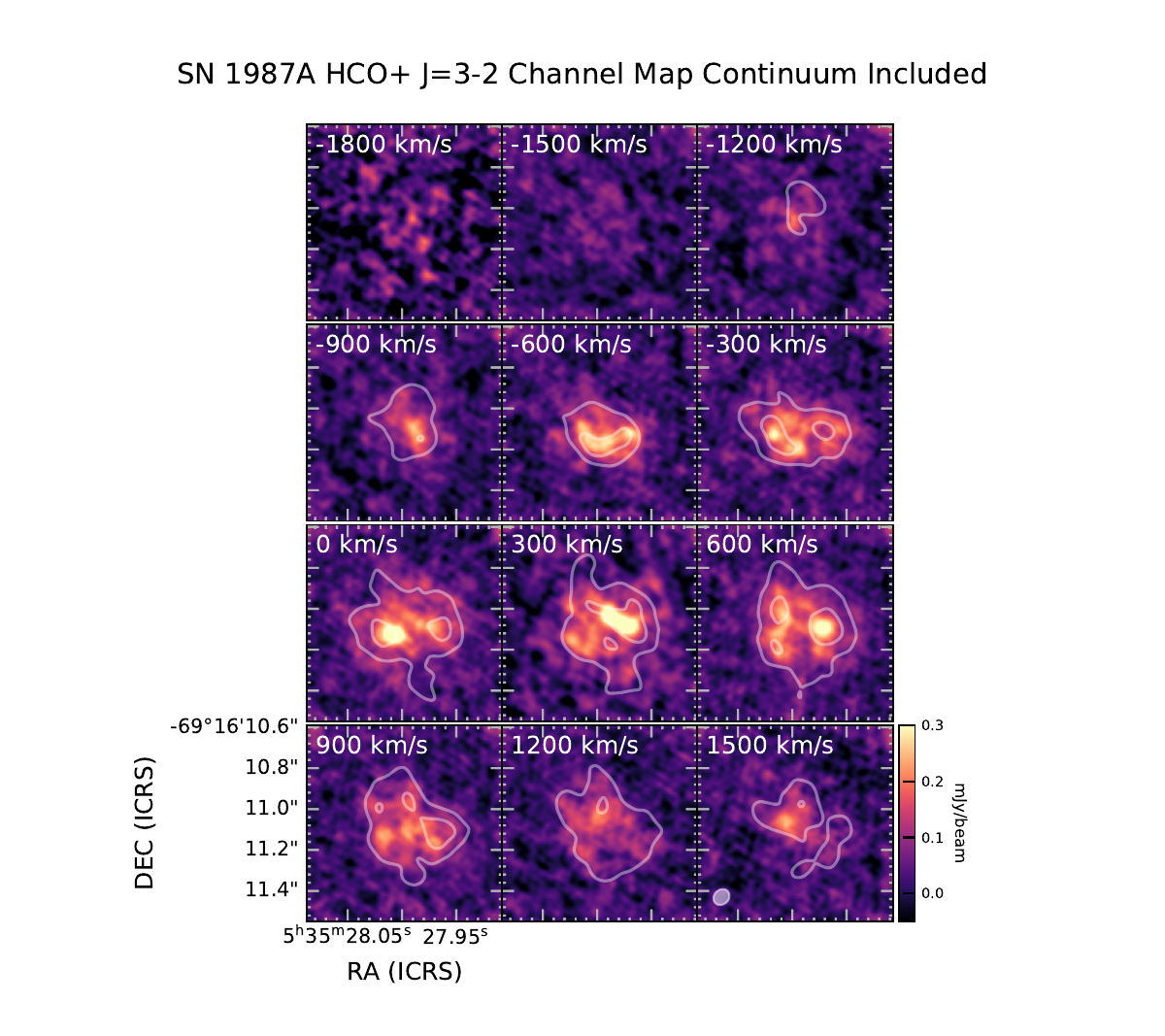}
    \caption{Velocity channel map of the $J=3-2$ transition of HCO$^+$ emission from velocities beginning at $-$1950\,km\,s$^{-1}$ to 1650\,km\,s$^{-1}$, binned in 300\,km\,s$^{-1}$ increments. Each image channel is labelled with the central velocity of the 300\,km\,s$^{-1}$ bin. Velocities are in LSRK.
    Continuum is included in this figure which, if uniformly spread out over the region, is $\sim$ 0.02\,mJy\,beam$^{-1}$, i.e. very low level.
    The typical RMS noise on these images is $\sim 2.5-4.3 \times 10^{-2}$\,mJy\,beam$^{-1}$.  The distribution of CO $J=2-1$ is superimposed as white line contours. The contour levels correspond to the CO brightness levels of 0.5 and 1.20\,mJy\,beam$^{-1}$. These levels were chosen to best represent the CO spatial distribution in relation to HCO$^+$. The spatial distributions of the CO and HCO$^+$ distributions are similar and peaks of high emission from the two species are at roughly the same locations, particularly in velocity channel 0\,km\,s$^{-1}$ to 600\,km\,s$^{-1}$. Note that all enclosed CO contours are bright spots in the CO distribution, except for the small ellipse in the central-south of channel 300\,km\,s$^{-1}$ where it is a hole. See \citet{Cigan_ref} for more details.}
    \label{fig:HCO+_4_fig}
\end{figure*}

%\mm{$-$1800\,km\,s$^{-1}$ to 1500\,km\,s$^{-1}$ (depending on how the bin is placed within 300\,km\,s$^{-1}$ grid (beginning or middle of the bin), this panel can be either "$-1800$\,km\,s$^{-1}$ to 1800\,km\,s$^{-1}$" or "$-$1950\,km\,s$^{-1}$ to 1650\,km\,s$^{-1}$"},

In order to form HCO$^+$, chemical reactions involving CO may be required, such as H$_3^+$ + CO $\rightarrow$ HCO$^+$ + H$_2$ \citep{mol_cloud, panessa_23} which will be discussed further in Sect.~\ref{disc:chemistry}. We analyse the spatially-resolved velocity channel maps of HCO$^+$ and CO emission and compare them to test this possibility, by examining whether the spatial distributions of CO and HCO$^+$ coincide or are similar. We would expect that their distributions would differ from that of SiO, because SiO is unrelated to the chemical reactions forming HCO$^+$.

The velocity channel map for the $J=3-2$ HCO$^+$ transition is displayed in colour in Fig.~\ref{fig:HCO+_4_fig}. A velocity channel map represents the line of sight Doppler shift caused by the expansion of the remnant, while the location of the emission shows dynamical motion of the gas since the SN explosion as projected on sky. Fig.~\ref{fig:HCO+_4_fig} is composed of 12 velocity channels from $-$1950\,km\,s$^{-1}$ to 1650\,km\,s$^{-1}$ and are binned in velocity increments of 300\,km\,s$^{-1}$. Each label on the velocity channels corresponds to the middle velocity of each bin. Bins of 300\,km\,s$^{-1}$ were chosen to maximise the signal in each bin whilst retaining velocity structure. These velocities are relative to the rest frequency of HCO$^+$ at the $J=3-2$ transition and comply with the Kinematic Local Standard of Rest frame (LSRK) used by ALMA. The systemic velocity of SN\,1987A of $v_{\rm helio}=286.7$\,km\,s$^{-1}$ ($v_{\rm LSRK}=274.2$\,km\,s$^{-1}$) \citep{Sys_vel_1987A_2008} has not been subtracted from the channel map. This is to aid our comparison of $J=3-2$ HCO$^+$ emission with the $J=2-1$ CO emission from \citet{Cigan_ref} which also has not been corrected for SN 1987A's systemic velocity.

The HCO$^+$ emission in Fig.~\ref{fig:HCO+_4_fig} starts appearing from velocities around $-$1200\,km\,s$^{-1}$, with the strongest emission found between channels at 0 and 600\,km\,s$^{-1}$, and fading towards 1500\,km\,s$^{-1}$. The brightest point is found in the west of the ejecta at the 300\,km\,s$^{-1}$ channel. The emission between the channels $-$600 to 300\,km\,s$^{-1}$ extends in the east and west direction, while the fainter emission spreads more towards the north and south direction, in particular at channels 600 to 1200\,km\,s$^{-1}$. 

 We also compare the spatial distribution of the CO emission in Fig.~\ref{fig:HCO+_4_fig}.
Line contours show the CO $J=2-1$ intensity spatial distribution from \citet{Cigan_ref}, being superimposed onto the HCO$^+$ spatial distributions at their corresponding velocity increments. The CO contour levels correspond to CO regions of brightness of 1.20 and 0.5\,mJy\,beam$^{-1}$ and were chosen to best-represent CO's regions of brightest intensity -- `bright spots' and the extended regions of CO. The CO line emission is, on average, $\sim$\,6.3 times brighter than the HCO$^+$ line emission.
Both intensity distributions show similarities with each other, namely the most intense regions in the central velocity channels $-$300\,km\,s$^{-1}$ to 600\,km\,s$^{-1}$ overlap. %\textst{There does however, appear to be a discrepancy between the weaker CO emission with}HCO$^+$ \textst{where there is an extension of the CO emission in the north-south direction in velocity channels 0 to 600}\,km\,s$^{-1}$. \textst{We can attribute this to the better sensitivity that the CO image has, which is able to capture the weaker flux emission} \citep{Cigan_ref}.

In comparison, there is little SiO emission spatially coincident with HCO$^+$ aside from bright spots in the 300 and 600\,km\,s$^{-1}$ velocity channels as seen in Appendix~\ref{app:sio} (Fig.~\ref{appendix:SiO_figure}). 
%Only one such similarity was uncovered when the same process was applied to compare HCO$^+$ with SiO \citep{Cigan_ref}, where the channels corresponding to 300 and 600\,km\,s$^{-1}$, some bright spots of HCO$^+$ overlapped with SiO. 

To further evaluate the similarities of the spatial distributions between HCO$^+$ with CO and SiO, a Spearman rank correlation function was used. The channel maps of the spatially-resolved HCO$^+$, CO and SiO images were convolved to have the beam size of the HCO$^+$ channel map using {\sc imsmooth}, a function of {\sc CASA}. To reduce the effect of oversampling, the pixel scale for all of the channels was increased to $\sim$0.036$\arcsec$ which roughly samples the FWHM beam twice using {\sc CASA}'s {\sc imrebin} function. The cross-correlation was measured between the brightness values of HCO$^+$ and CO pixels in the velocity channels ranging from $-$600 to 600\,km\,s$^{-1}$, i.e. the five brightest channels in the maps. The effect of the continuum-level present in the HCO$^+$ images was reduced by setting a brightness threshold of 3\,$\times$\,RMS in each channel (roughly 0.07$-$0.1 mJy\,beam$^{-1}$) to omit low-brightness background pixels. The Spearman correlation test statistic, for HCO$^+$ versus CO is 0.72, %0.74, 
indicating a moderate correlation of morphologies between the two molecules. 

From Sect.~\ref{sect:Continuum_estimation}, the spatial distribution of the dust emission peaks in the location of low HCO$^+$ intensity, however, there is still low-level dust emission overlapping with the HCO$^+$ emission. Therefore, we further investigate whether these dust levels from our best-fitting modified blackbody at 267\,GHz impact the continuum included $J=3-2$ HCO$^+$ spatial distributions. We scale the continuum levels by the beam of the HCO$^+$ $J=3-2$ emission which leads to a uniform continuum level intensity of 0.02\,mJy\,beam$^{-1}$ spread across the ejecta. Since this is $\sim$4 times smaller than our RMS threshold, the continuum levels have a minimal impact on the correlation analysis.
%The spatial similarities between the two distributions are predominantly between CO and HCO$^+$ as we assume that the continuum levels are due to dust which was shown to be uncorrelated with the $J=2-1$ transition of CO \citep{Cigan_ref}, there could be a contribution from the continuum in the form of higher brightness clumps which cannot entirely be ruled out however. 
%\hg{rephrase so that you let reader know you have changed the pixel size and took a SN threshold to account for continuum and then remove the rest}

This correlation was performed through use of {\tt SciPy}'s {\sc stats.spearmanr} module\footnote{\url{https://docs.scipy.org/doc/scipy/reference/generated/scipy.stats.spearmanr.html}}. The same process was applied to test the correlation between HCO$^+$ and SiO, which returned a weaker correlation strength of %0.67
 0.65 compared to HCO$^+$ and CO. The HCO$^+$ versus SiO correlation is still strong however and a possible explanation for this is that it could be showing a strong correlation for the ejecta rather than the individual regions of brightness within the ejecta, the same can be said for CO. Despite this, we can still trust the spatial distributions of HCO$^+$ compared to CO and SiO in Fig.~\ref{fig:HCO+_4_fig} and ~\ref{appendix:SiO_figure} respectively which show far more similarities between distributions of CO and HCO$^+$ compared to similarities between SiO and HCO$^+$.

A velocity channel map of HCO$^+$ at the higher transition of $J=4-3$ is provided in Fig.~\ref{appendix:HCO_43}. There is also emission from the continuum present in this channel map. The $J=4-3$ HCO$^+$ and continuum emission in the ejecta is only marginally spatially resolved, while synchrotron emission from the ring is detected across the channels.
Due to limitations in frequency coverage at the higher transition of HCO$^+$ at $J=4-3$, the channel map only spans from $-$1350\,km\,s$^{-1}$ to 1650\,km\,s$^{-1}$ respective to the rest frequency of HCO$^+$ at the $J=4-3$ transition, thus we are missing the corresponding $-$1800 and $-$1500 \,km\,s$^{-1}$ binned velocity channels which are present in the $J=3-2$ HCO$^+$ transition in Fig.~\ref{fig:HCO+_4_fig}. 

\subsection{Comparison of H$\alpha$ with HCO$^+$}\label{H_alpha_comp}
\begin{figure*}
    \includegraphics[width=18cm]{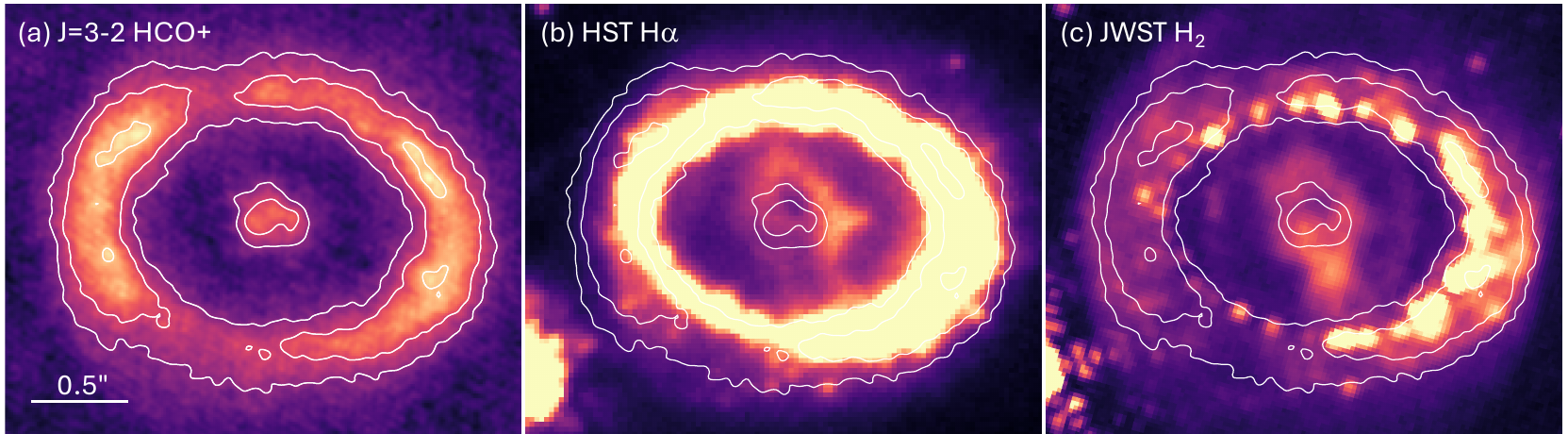}
    %\caption{ The {\it HST} WFC3 F625W image, which predominantly contains H$\alpha$ emission, shows the `keyhole' shape of the ejecta of SN 1987A. The ring also shows emission from H$\alpha$ which is over-saturated due to surface brightness boundaries chosen to best display the H$\alpha$ distribution in the ejecta. The white contours show the continuum included $J=3-2$ HCO$^+$ emission obtained by integrating across velocity space. The contour level of the HCO$^+$ emission is 0.3 and 0.5 Jy\,beam$^{-1}$ km\,s$^{-1}$. Although both H$\alpha$ and HCO$^+$ peak in the west side of the ejecta, the remainder of HCO$^+$ emission does not have strong correspondence with the H$\alpha$ emission.}
    \caption{Comparison of (a) ALMA $J=3$--$2$ HCO$^+$, (b) {\it HST} H$\alpha$, and (c) {\it JWST} H$_2$ emission. (a) Continuum-included $J=3$--$2$ HCO$^+$ image, obtained by integrating over velocity space. The white contours indicate HCO$^+$ emission levels of 0.3, 0.5 and 0.85 Jy\,beam$^{-1}$\,km\,s$^{-1}$ in all three panels. 
     The emission from the ring is due to synchrotron radiation, not HCO$^+$ emission.
    (b) {\it HST} WFC3 F625W image, which predominantly traces H$\alpha$ emission, revealing the characteristic `keyhole' morphology of the SN~1987A ejecta. (c) {\it JWST} F212N image, primarily tracing the 2.12~$\mu$m H$_2$ emission. The peak of the HCO$^+$ emission lies along the edge of the cavity in the H$\alpha$ `keyhole' structure. The remaining HCO$^+$ emission does not have strong correspondence with either the H$\alpha$ or H$_2$ emission.}
    \label{fig:H_alpha_HCO}
\end{figure*}

%Checked plot, contour originates from 9 channel v-chan map so we're good :) 
To investigate whether HCO$^+$ forms due to ionisation of the ejecta gas caused by X-ray energy deposition from the ring, we compare the $J=3-2$ HCO$^+$ spatial distribution with that of H$\alpha$ emission in the ejecta. Fig.~\ref{fig:H_alpha_HCO} shows the comparison between the {\it HST} F625W band image (dominated by H$\alpha$ emission) and the $J=3-2$ HCO$^+$ distribution (white contour). The HCO$^+$ contour is the result of integrating the HCO$^+$ $J=3-2$ velocity channel map (seen in Fig.~\ref{fig:HCO+_4_fig}) across the velocity range $-$1050 to 1650\,km\,s$^{-1}$.
Comparing the two distributions, we see that the HCO$^+$ distribution is more compact than the H$\alpha$ emission, and it's located to the north of the overall keyhole-shaped H$\alpha$ morphology. This compactness of the HCO$^+$ emission compared to H$\alpha$ indicates that HCO$^+$ does not form in the strongly-ionised outer regions of the ejecta, but instead may form deeper within the ejecta, where the gas is only mildly ionised. Additionally, we don't expect HCO$^+$ to exist in the more strongly ionised regions of the ejecta, as X-rays can dissociate molecules and therefore prevent HCO$^+$ from forming.
%\hg{has keyhole been defined before eg in intro? if not, it will need to be defined somewhere} 
It is apparent that the region of highest HCO$^+$ intensity (the north contour) coincides with the dip region of brightest H$\alpha$ emission, and at the centre of the system \citet{Larsson_2016} . This was further investigated by comparing the line velocity profiles of H$\alpha$ found in \citet{Larsson_2016} with that of HCO$^+$. In the rest frame, the majority of the H$\alpha$ emission is blue-shifted in the central region of the ejecta while the majority of the central HCO$^+$ emission exists at low red-shifts as seen in Fig.~\ref{fig:HCO+_4_fig}. This means that the HCO$^+$ and H$\alpha$ emission are not co-spatial with each other in velocity space, with HCO$^+$ residing just behind the H$\alpha$ emission in the ejecta. It is difficult however, to entirely rule out that HCO$^+$ forms from the ionisation of molecules via deposition of X-ray energy; the south-eastern and north parts of the HCO$^+$ contour lies in a region of little H$\alpha$ emission. This could be partly because of H$\alpha$ emission in this region being obscured by dust \citep{Cigan_ref, Matsuura_2024_H2}. 

\subsection{Measurements of Line Intensities} \label{section:LE_measurements}

\begin{table*}
\begin{tabular}{ |p{2.0cm}||p{2.0cm}|p{2.0cm}|p{2.0cm} | p{2.0cm}|p{2.0cm}|p{2.0cm}|p{2.0cm}}
 \hline
  Transition & $I_{\rm tot}$ [$10^{-20}\,\rm W/m^2$]  & $C_{\rm dust}$ [$10^{-20}\,\rm W/m^2$] & $I_{\rm HCO^+}$ [$10^{-20}\,\rm W/m^2$] & Error$^a$ [$10^{-20}\,\rm W/m^2$]& Error$^b$ [$10^{-20}\,\rm W/m^2$] & Error$^c$ [$10^{-20}\,\rm W/m^2$]\\
 \hline
$J=3-2$ &10.52 &2.69 &7.83 &$\pm$\,0.55 &$\pm$\,0.18 &$\pm$\,0.78 \\
$J=4-3$  &15.87 &10.17 &5.69 &$\pm$\,0.40 &$\pm$\,0.24 
&$\pm$\,2.54 \\ 
\hline
% Dust range J=4-3 = {$\pm _{2.436}^{3.219}$}
\end{tabular}

\caption{Line intensities and continuum estimations for HCO$^+$. $I_{\rm tot}$ - the total flux for the respective transition, this includes the line intensity from HCO$^+$ and the intensity level from the continuum.  $C_{\rm dust} $ - the dust continuum levels estimated using the fitted dust SED to observations in Figure~\ref{fig:therm_dust}. $I_{\rm HCO^+}$ - the line intensity for HCO$^+$ with the continuum levels subtracted. Error $^a$ - Uncertainty of flux calibration in ALMA data cubes. Error $^b$ - RMS in data cubes. Error $^c$ - Uncertainty in continuum level estimation (30$\%$ of $C_{\rm dust}$). \label{tab:HCO_fluxes}}
\end{table*}
%Flux Density plots with Continuum level
\begin{figure}
	\includegraphics[clip, trim=1.2cm 0.5cm 1.5cm 2.0cm, width=\columnwidth]{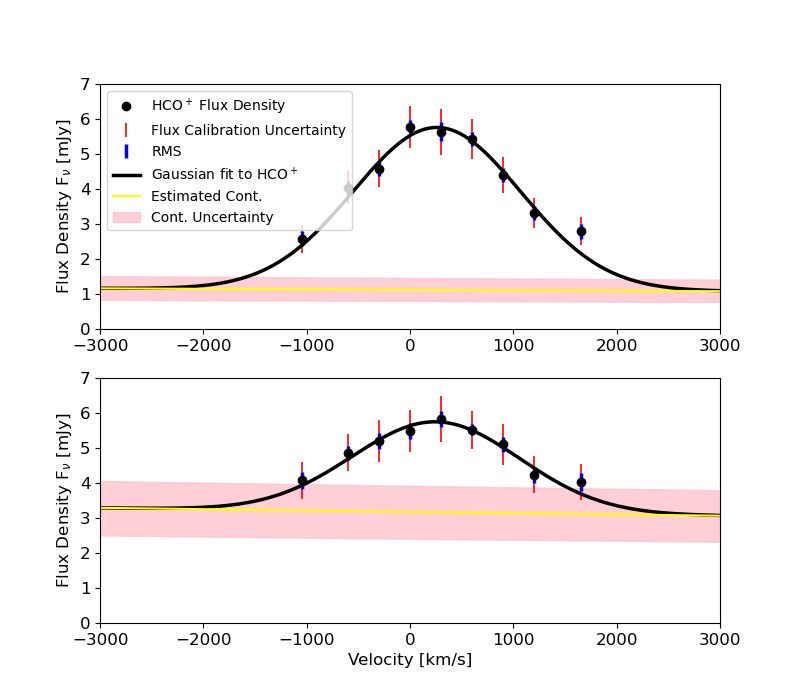}
    \caption{Upper: Flux density profile for the $J=3-2$ HCO$^+$ with continuum included in the flux density. This profile was fitted with a gaussian curve to retrieve the FWHM of the line which was calculated to be $+$1906\,km\,s$^{-1}$ shown as a black line.
    Lower: Flux density profile for the $J=4-3$ HCO$^+$ transition with continuum included and fitted with a Gaussian curve with FWHM fixed at 1906\,km\,s$^{-1}$ (black line). Both transitions include uncertainties of the RMS and band calibration shown as blue and red error bars. These figures also show the best fit line for the continuum level (yellow line) and its uncertainty of fit (pink shaded region), obtained from a Monte-Carlo simulation fitting thermal dust emission by a modified blackbody spectrum.} %\mm{In the labels, 'MC Best fit' and 'Fit uncertainty' are actually,'(estimated) Continuum' and 'continuum uncertainty'? (Fit can be Gaussian fit in this figure, so confusing). You can shorten as 'cont.' to save space}
    \label{fig:FD_plots}
\end{figure}

To estimate the physical properties (temperature and mass) of HCO$^+$ emission within the ejecta, we first extract the intensity from the channel maps to create the flux profile of HCO$^+$, then integrate across the frequency space to obtain the line intensity. Line intensities can later be compared to modelled line intensities using the radiative transfer code {\sc RADEX} to estimate the temperature and mass, details of which are found in Section~\ref{RADEX}. The continuum levels present are also estimated and subtracted from the observed line intensities to ensure emission is mostly HCO$^+$. The flux profile was extracted from the HCO$^+$ $J=3-2$ continuum-included channel map using aperture photometry within {\sc CASA}, specifically using the {\sc image.getprofile} function, where beam sizes, velocity increments and aperture size across frequency were kept constant. The aperture was an ellipse of semi-major and semi-minor axes of 0.368${\arcsec}$\,$\times$\,0.446${\arcsec}$ respectively with an inclination angle of 0$^\circ$. 
%\mm{ (FWHM? or the cut-off radius?) } \hd{This is a cut-off radius, I've decided not to alter it in text as I think me specifying its an aperture not a beam size makes the reader think it is an area of those dimensions. Specifying that it isn't a FWHM would only confuse the reader.}
This aperture was chosen to capture the most of the HCO$^+$ emission.
As seen in Figure~\ref{fig:HCO+_4_fig} the velocity channels corresponding to $-$1800, $-$1500 and $-$1200\,km\,s$^{-1}$ have negligible\footnote{The flux measured within the ejecta in the $-$1800\,km\,s$^{-1}$ to $-$1200\,km\,s$^{-1}$ channels may make up only 9$-$12\% of the total HCO$^+$ flux from all channels which is within the uncertainty level.} HCO$^+$ emission. After realising that these channels add more noise than signal, these channels were omitted from the measurement of the $J=3-2$ HCO$^+$ line intensity. 

The $J=4-3$ HCO$^+$ emission was analysed across the same velocity range as the $J=3-2$ HCO$^+$ emission despite there being a detection of possible emission in the $-$1200\,km\,s$^{-1}$. We omit this channel from analysis due to there being an insufficient detection of HCO$^+$ emission in the corresponding $-$1200\,km\,s$^{-1}$ channel of the $J=3-2$ HCO$^+$ emission. Additionally, it might be possible that the ejecta emission at the $-$1200\,km\,s$^{-1}$ channel in Figure~\ref{appendix:HCO_43} is due to dust emission and not from HCO$^+$ $J=4-3$ line emission.

%\mm{I am not sure if 'line emission' is the correct term for this context. Line emission refers to the line to be seen as emission, opposed to absorption, but it does not carry any physical/energy unit (flux or intensity). Better to use 'line intensity', here. Sometimes, I have seen 'emission intensity' or '(emission) line strength'. In  Goldsmith \& Langer (1999), bible of the population diagram, 'integrated line intensity'.}
The line intensity of HCO$^+$ was calculated by integrating the continuum-included HCO$^+$ fluxes across the $J=3-2$ and $J=4-3$ frequency ranges. 
The best-fitting modified blackbody curve as seen in Figure~\ref{fig:therm_dust} and also in Figure~\ref{fig:FD_plots}, was integrated across the HCO$^+$ frequency ranges and subtracted from the continuum-included HCO$^+$ line intensity. 

We consider mainly three contributions of uncertainties to the HCO$^+$ line intensities: the RMS in the channel map, the ALMA band calibration uncertainty (otherwise known as the flux calibration uncertainty) and the uncertainty in the estimated continuum level.
The largest source of error in the HCO$^+$ line intensities for both the $J=3-2$ and $J=4-3$ transitions is the uncertainty of the underlying continuum (Table~\ref{tab:HCO_fluxes}) which is 7\,$\%$ and 300\,$\%$ larger than the combined flux calibration uncertainties and RMS for the $J=3-2$ and $J=4-3$ transitions respectively. 

%\mm{this is larger than RMS noise (the last column of the table) and flux calibration uncertainties (***\%\, give values in ***).} Errors in continuum level estimation originate from the errors in the fit of the free parameters. 
%Since we have two different methods of defining the continuum for the $J=3-2$ emission: adjacent ALMA spectral window analysis and the LMC dust, we quote the difference of continuum levels of these two methods, as the error for the line intensity of the $J=3-2$ HCO$^+$ emission. 
To visualise how the estimated continuum level compares with the HCO$^+$ flux profiles, Figure~\ref{fig:FD_plots} shows the flux profiles for HCO$^+$ at the $J=3-2$ and $J=4-3$ transitions. The estimated continuum level and uncertainties from the MC simulated modified blackbody fit is shown as the solid yellow line and pink shaded region respectively. Complete spectral coverage, out beyond $\pm$\,2000\,km\,s$^{-1}$, of the line profile was not achieved and hence direct measurement of the flux observations underestimates the total line emission from HCO$^+$. Properties such as the FWHM of the flux density profiles were needed for analysis in Sect.~\ref{RADEX} and so we modelled a gaussian fit to our continuum-subtracted observations and extrapolate to reach HCO$^+$ fluxes around 0\,Jy. The resulting fit is shown by the solid black line in Figure~\ref{fig:FD_plots}, note that the systematic continuum levels have been added back on. Errors in the calibration of the flux, RMS and uncertainties from the continuum level estimations were included in the fit error and whose values are seen in Table~\ref{tab:HCO_fluxes}. Fitting the $J=3-2$ HCO$^+$ line with a gaussian model resulted in a FWHM of 1906\,km\,s\,$^{-1}$ which was fixed at this value to also fit the $J=4-3$ HCO$^+$ transition. The symmetrical center of the gaussian fit to the flux profiles are 260\,$\pm$\,60\,km\,s$^{-1}$ and 250\,$\pm$\,200\,km\,s$^{-1}$ for the $J=3-2$ and $J=4-3$ HCO$^+$ transitions respectively. These are consistent with the systemic regressional velocity of SN 1987A of $v_{\rm helio}=286.7$\,km\,s$^{-1}$ ($v_{\rm LSRK}=274.2$\,km\,s$^{-1}$) \citep{Sys_vel_1987A_2008}.

The continuum-subtracted line intensity of the HCO$^+$  $J=3-2$ and the $J=4-3$ transitions are therefore:  $I_{\rm HCO+}$ = (7.8 $\pm$ 0.5) $\times 10^{-20}$ W\,m$^{-2}$ and  (5.7 $\pm$ 1.0) $\times 10^{-20}$ W\,m$^{-2}$ (Table~\ref{tab:HCO_fluxes}). The HCO$^+$ $J=3-2$ line intensity is in agreement, within 3$\sigma$, with a previously published measurement of (6.7$\pm$ 0.6)$\times$10$^{-20}$ W\,m$^{-2}$ based on the spectral line survey \citep{Mikako_ref} with our measured line intensity being larger. 

%%%%%%%%%%%%%%%%%%%%%%%%%%%%%%%%%%%%%%%%%%%%%%%%%%%%%%%%%%%%%%%%%
\section{HCO$^+$ Line Intensity Analysis with non-LTE code}
\label{RADEX}

% Chi square plot
\begin{figure*}
	\includegraphics[clip, trim=2.8cm 0.1cm 1.5cm 0.1cm, width=15cm]{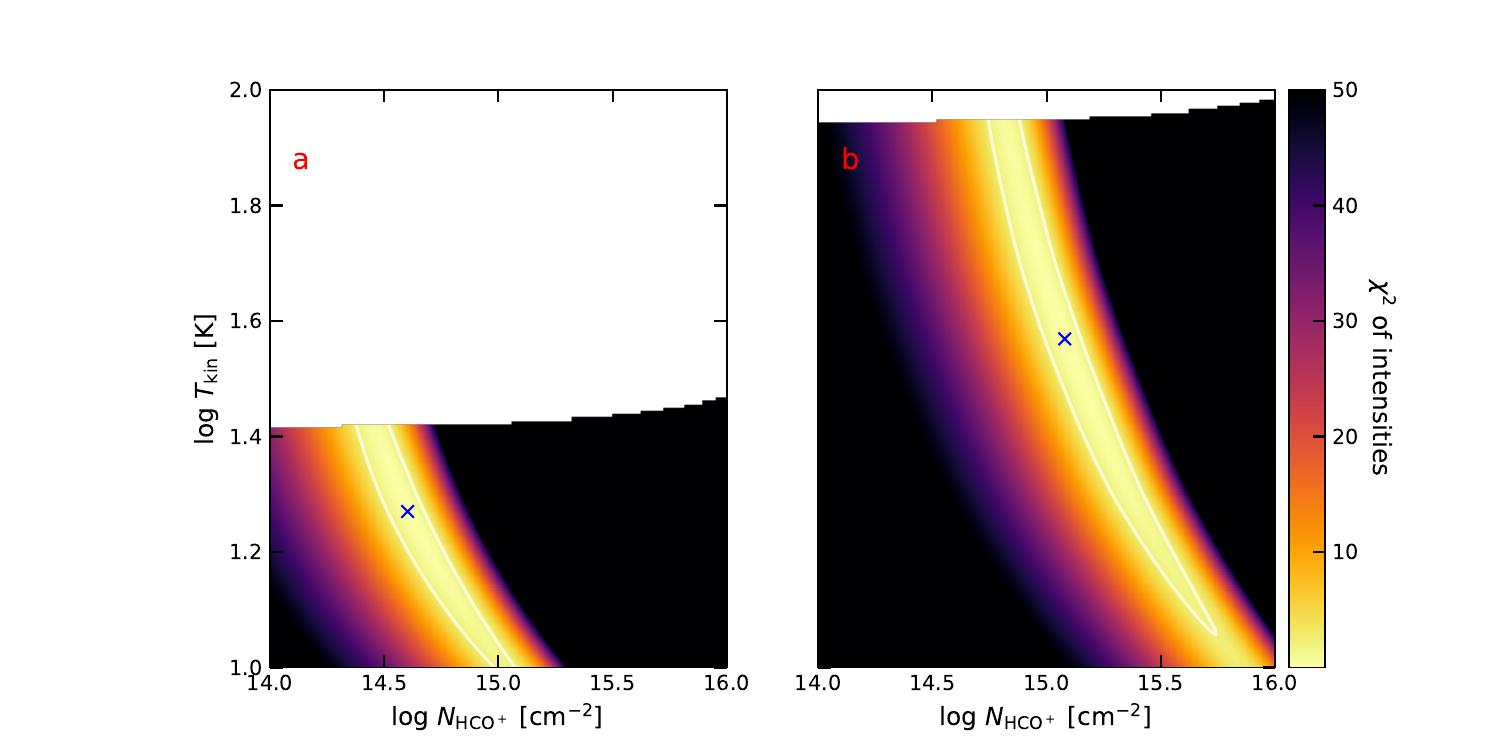}
    \caption{$\chi^2$ distribution for comparisons of {\sc RADEX} line intensities with observed line fluxes with respect to the {\sc RADEX} input parameters of kinetic temperature, $T_{\rm kin}$, and column density, $N_{\rm HCO^+}$, of HCO$^+$. ($a$) and ($b$) show the results using the continuum-subtracted HCO$^+$ line intensity at a H$_2$ collisional partner density of at 1$\times10^{6}$ and 1$\times10^{5}$\,cm$^{-3}$ respectively. The white space in the figures indicate the negative $\tau$ encountered while running {\sc RADEX} and hence resulting $\chi^2$ values are omitted from further analysis. The white contour represents the 1$\sigma$ standard deviation of the $\chi^2$ distribution. The blue cross marks the position of the minimum $\chi^2$ value and therefore the best-fit parameters for $N_{\rm HCO^+}$ and $T_{\rm kin}$.}
    \label{fig:chi_square}
\end{figure*}

\subsection{RADEX}\label{Sect:Radex}
To infer the mass and temperature of HCO$^+$ from the measured line intensities, comparisons were made with 
theoretically calculated line intensities from a non-LTE (non-local thermodynamic equilibrium) radiative transfer code. 
We used the python version\footnote{\url{ https://pythonradex.readthedocs.io/en/latest/}} of {\sc RADEX} \citep{RADEX}. {\sc RADEX} requires the input of parameters including kinetic temperature ($T_{\rm kin}$) and HCO$^+$ column density ($N_{\rm HCO^+}$): by fine tuning these parameters and replicating the measured intensities via $\chi^2$ analysis, we will obtain $T_{\rm kin}$ and $N_{\rm HCO^+}$. The range of $T_{\rm kin}$ was varied from 10 to 100\,K and $N_{\rm HCO^+}$ ranged from 1$\times$10$^{14}$ to 1$\times$10$^{16}$\,cm$^{-2}$, both were varied at logarithmic interval steps. 

Molecular constants, such as Einstein coefficients, collisional rates, transitions and corresponding statistical weights for HCO$^+$ were obtained from the {\sc LAMDA} database \citep{LAMDA}. Physical properties such as velocity line-widths and solid angle of the HCO$^+$ emission were taken from our ALMA observations. We adopted a FWHM line-width of 1906\,km\,s$^{-1}$, as found in Sect.\,\ref{section:LE_measurements} for HCO$^+$ $J=$3--2 line.
The solid angle was calculated using the elliptical aperture radii of flux extraction in the previous subsection: 0.368\,$\arcsec$ $\times$ 0.446\,$\arcsec$. The cloud geometry was assumed to be spherical. Additionally, to solve the radiative transfer equation, the collisional partner and its corresponding density had to be estimated. We assume the dominant collisional partner in the ejecta to be H$_2$ as only collisional information of H$_2$ with HCO$^+$ was available to us in {\sc LAMDA}. We estimate an approximate H$_2$ density by calculating it from the mass of the hydrogen envelope in SN 1987A, assumed to be to 6\,M$_\odot$ (e.g. \citealt{Jerkstrand}) and the volume of the ejecta at the end day of the HCO$^+$ observations. The volume of the ejecta is derived from a free expansion velocity of the ejecta to be 1300\,km\,s$^{-1}$, estimated from the FWHM line emission of HCO$^+$ line (1900\,km\,s$^{-1}$), and the relation between the FWHM and expansion velocity from \citet{McCray_1993}. Hence, the estimated H$_2$ density at the time of our observations is roughly 3$\times$10$^5$ cm$^{-3}$. 
Therefore, we fix our H$_2$ densities at 1$\times$10$^{6}$ and 1$\times$10$^{5}$\,cm$^{-3}$. We adopt the higher H$_2$ collisional partner density to aid with our comparison of HCO$^+$ masses with those found in \citep{Mikako_ref} who use a H$_2$ collisional partner density of 10$^6$\,cm$^{-3}$.
Additionally, these H$_2$ densities are also consistent with those derived by \citet{Larsson_model_2023} based on \textit{JWST} observations of H$_2$ in SN 1987A, these densities can only be achieved however, if a H$_2$ clumping factor is significant.
%The H$_2$ density was fixed at 1$\times$10$^{6}$ and 1$\times$10$^{5}$\,cm$^{-3}$. We believe these densities are reasonable by extrapolating O/C densities at 8 years after the SN explosion, modelled in \citet{Jerkstrand}, to the time of the HCO$^+$ observations at 30$-$32 years since the SN explosion.
%\textcolor{magenta}{These H$_2$ densities are also consistent with those derived by \citet{Larsson_model_2023} based on \textit{JWST} observations of H$_2$ in SN 1987A, these densities can only be achieved however, if a H$_2$ clumping factor is significant.}
%Assuming a free expansion velocity of the ejecta to be 1300\,km\,s$^{-1}$, estimated from the FWHM line emission of HCO$^+$ line (1900\,km\,s$^{-1}$), and the relation between the FWHM and expansion velocity from \citet{McCray_1993}, the estimated O/C density at the time of our observations is roughly 10$^5$ cm$^{-3}$. 
%\hd{Assuming a uniform sphere of HCO$^+$ emission}

In order to find the input parameters that return the best-matching line intensities to our HCO$^+$ line intensities, a $\chi^2$ goodness-of-fit test was set up where $\chi^2_{\rm min}$ is the best fit of {\sc RADEX} line intensities to the observed line intensities. Figure~\ref{fig:chi_square} shows the summed $\chi^2$ distribution for the two HCO$^+$ lines $J=3-2$ and $J=4-3$. Subplots (a) and (b) on Figure~\ref{fig:chi_square} display the $\chi^2$ distributions where the H$_2$ collisional partner density is fixed at 1$\times$10$^6$\,cm$^{-3}$ and at 1$\times$10$^5$\,cm$^{-3}$ respectively and the positions of $\chi^2_{\rm min}$, the best-matching line intensities to the observed HCO$^+$ line intensities, are marked with a blue cross. These $\chi^2_{\rm min}$ values are 0.4$\times$10$^{-3}$ for (a) and 0.6$\times$10$^{-3}$ for (b). A confidence interval in the $\chi^2$ space was also set up where a $\Delta\chi^2$ of 2.30 was chosen to account for the variation of two interesting parameters ($N_{\rm HCO^+}$ and $T_{\rm kin}$) at a 1$\sigma$ significance level, this is displayed as a white ellipse on Figure~\ref{fig:chi_square}. Unfortunately, $T_{\rm kin}$ and $N_{\rm HCO^+}$ are fairly degenerate which originates from the large uncertainties in the estimated continuum level. The corresponding best fit values for HCO$^+$ column density and kinetic temperature are therefore:
$N_{\rm HCO^+}=$(4.00 $ \,^{+7.8}_{-1.6}$)$\times$10$^{14}$\,cm$^{-2}$ and $T_{\rm kin}$=18.6$\,^{+7.6}_{-8.6}$\,K for a H$_2$ collision partner density of 10$^6$\,cm$^{-3}$ and $N_{\rm HCO^+} =$(1.20 $ \,^{+4.3}_{-0.6}$)$\times$10$^{15}$\,cm$^{-2}$ and $T_{\rm kin}$=37.1$\,^{+51.7}_{-25.6}$\,K for a H$_2$ collision partner density of 10$^5$\,cm$^{-3}$.

There was a limitation in our input parameter ranges and $\chi^2$-analysis. {\sc RADEX} returns a negative optical depth (hereafter $\tau$) value for the HCO$^+$ $J=1-0$ transition at high temperatures (beginning at $\sim$26\,K and $\sim$89\,K for 10$^6$ and 10$^5$ cm$^{-3}$ H$_2$ collisional partner densities respectively). A negative value for $\tau$ corresponds to a numerical error rather than an actual solution \citep{RADEX}.%, as {\sc RADEX} does incorporate population inversion (e.g. masers). 
Therefore all {\sc RADEX} outputs containing negative $\tau$ have been excluded from our analysis.

\subsection{The Mass of HCO$^+$}
The mass of HCO$^+$ can be calculated from the column density through the use of the following equation: $M_{\rm HCO^+} = f\Omega  N_{\rm HCO^+}d^2m$ where $M_{\rm HCO^+}$ is the mass of HCO$^+$ in kg, $f$ is the filling factor (assumed to be 1 for this study), $\Omega$ is the solid angle in steradians, $N_{\rm HCO^+}$ is the column density of HCO$^+$ in cm$^{-2}$, $d$ is the distance to SN 1987A from Earth in cm ($d$ taken here to be 51.2\,kpc, \citealp{distance_1991}) and $m$ is the molecular mass of HCO$^+$. The uncertainty in the mass was propagated using the 1$\sigma$ confidence levels in $N_{\rm HCO^+}$ taken from the $\chi^2$ distribution. 
Using the equation above and $N_{\rm HCO^+}$ from the best-fit results of the $\chi^2$ analysis in the previous section, the mass of HCO$^+$ was found to be: (2.9 $\,^{+5.7}_{-1.2}$)$\times$10$^{-6}\,$M$_{\odot}$ for a hydrogen collisional partner density of 1$\times$10$^6$\,cm$^{-3}$ and (8.8$\,^{+31.5}_{-4.7}$)$\times$10$^{-6}\,$M$_{\odot}$ for a hydrogen collisional partner density of 1$\times$10$^5$\,cm$^{-3}$.

%\hd{I changed the of the following, before it argues that masses were much larger than Matsuura et al. but that is not entirely true, if gas densities are the same at 1e6 cm-3 then the masses are more or less consistent with eachother. Only our mass with a lower gas density is larger than Matsuura et al's mass. }
The mass of HCO$^+$ in SN 1987A was first estimated using the $J=3-2$  line intensity and an upper limit for the $J=4-3$ line intensity \citep{Mikako_ref}. They found an upper limit of $M\rm _{HCO^+} \lesssim 5\times 10^{-6} \,\rm M_{\odot}$. For the same H$_2$ collisional partner density of 1$\times$10$^{6}$\,cm$^{-3}$, our value of $M\rm _{HCO^+}$ at 3$\times$10$^{6}$\,M$_\odot$ is smaller but consistent within our uncertainties. Although slightly different $d$ and $\Omega$ values have been used in these two works, these differences are not the major contributors to the different HCO$^+$ masses: our distance is only $\sim$2\,$\%$ larger than the 50\,kpc used in \citeauthor{Mikako_ref}, and our solid angle is approximately a factor of two smaller. 
%Adopting our values $\Omega$ and $d$ in the mass calculation of \citeauthor{Mikako_ref} leads to a mass of 2.4$\times$10$^{-6}$\,M$_\odot$, resulting in a difference of only $\sim$2.6$\times$10$^{-6}$\,M$_\odot$ of HCO$^+$ mass. 

Adopting the lower collisional gas density of 1$\times$10$^5$\,cm$^{-3}$ could be more representative of typical densities within the remnant for our 2019 observations as $\sim$4 years have elapsed between the 2015 observations of \citet{Mikako_ref} and the end of our HCO$^+$ observations, where the ejecta expands in time and thus the gas density decreases. %\textcolor{magenta}{This was also supported in the earlier Sect.~\ref{Sect:Radex}.}
If the lower collisional partner gas density is assumed, then our calculated HCO$^+$ mass (9$\times$10$^{-6}$\,M$_\odot$) exceeds the HCO$^+$ mass found by \citet{Mikako_ref}.

%%%%%%%%%%%%%%%%%%%%%%%%%%%%%%%%%%%%%%%%%%%%%%%%%%%%%%%%%%%%%%%%%

\section{Discussion}
The detection of emission from two HCO$^+$ transitions at $J=3-2$ and $J=4-3$ confirms the presence of HCO$^+$ within the ejecta of SN 1987A. Due to HCO$^+$ being a novel detection in SN 1987A \citep{Mikako_ref}, there are few studies which predict its formation in supernova remnants. In the following section, the observations of HCO$^+$ are used to highlight potential chemical reactions which can form it, to infer how it is able to form in SN 1987A and to discuss what are the current conditions in ejecta regions where we find HCO$^+$.

\subsection{Chemistry} \label{disc:chemistry}

We roughly estimate the total amount of HCO$^+$ formed over the lifetime by considering only the dominant formation pathway as a feasibility test of whether the observed HCO$^+$ mass can be produced.
In the ISM, HCO$^+$ can form via the following chemical reaction:
\begin{equation}\label{eqn:HCO_form_react}
    \rm CO + H_3^+ \rightarrow HCO^+ + H_2
\end{equation}
which has a high reaction rate coefficient, $k(T)$ across a range of temperatures from 10$-$300\,K according to the {\sc UMIST} database for astrochemistry \citep{UMIST_2024, UMIST_95, UMIST_91, UMIST_12, UMIST_06, UMIST_99} which is reminiscent of the temperatures of the ejecta after $d_{\rm SN} \sim 1,000$. 
%\textcolor{magenta}{A higher $k(T)$ indicates that the reaction is likely to occur quickly within the ejecta and hence can form more HCO$^+$ within a select timescale compared to other reactions which may have a lower $k(T)$.}
We suggest that this reaction may be the primary HCO$^+$ formation reaction in the SN ejecta due to the similarity in the spatial distributions of our $J=3-2$ HCO$^+$ emission and $J=2-1$ CO emission and due to the high abundance of CO in the ejecta \citep[$\sim$0.02$-$1\,M$_\odot$; ][]{Mikako_ref}. We estimate the mass of HCO$^+$ formed over $\sim$30 years by using this chemical reaction rate. 

Additional HCO$^+$ formation reactions involving other reactants could also potentially occur such as CH$^+$+H$_2$O $\rightarrow$ HCO$^+$ + H$_2$ and CO$^+$ + HCO $\rightarrow$ HCO$^+$ + CO as examples. For the reasons above and the fact that molecules like CH$^+$ and H$_2$O have yet to be detected in SN 1987A, we therefore just consider the reaction in Eq.~\ref{eqn:HCO_form_react} which has fewer unknown parameters.

Firstly, we set an appropriate timescale over which Eq.~\ref{eqn:HCO_form_react} occurs. This is between $d_{\rm SN}$ = 1,000 and 11,499. The start day of $d_{\rm SN}$=1,000 was chosen because it corresponds to when masses of CO and H$_2$ in chemical models become stable in the ejecta \citep{Culhane_McCray_H2_1995, sarangi_cherchneff, 2018MNRAS.480.5580S, Ono_matter_mixing_2024}. $d_{\rm SN} = 11,499$ corresponds to the median day of our HCO$^+$ observations. 

The rate of HCO$^+$ formation, $R{\rm_{HCO^+}}$, for Eq.~\ref{eqn:HCO_form_react} is defined as $R{\rm_{HCO^+}} = k(T)n_{\rm CO}n_{\rm H^+_3}$ which depends on the number densities, $n$, of the reactants CO and H$_3^+$ and also $k(T)$ which is the reaction-rate coefficient for Eq.~\ref{eqn:HCO_form_react}. For simplicity, we assume a uniform distribution of CO and H$_3^+$ molecules across the ejecta; i.e. the ejecta gas is chemically well-mixed {\bf and the number densities are constant across}. $n_{\rm CO}$ is calculated from the CO masses in \citet{Mikako_ref}. We approximate $n_{\rm H^+_3}$ by its relationship in dense clouds in the ISM as $n_{\rm H^+_3} = \dfrac{\zeta}{k_{\rm H^+_3}(T)}\times \dfrac{n_{\rm H_2}}{n_{\rm CO}}$ \citep{Oka_H3_06}. $\zeta$ is the ionisation rate of H$_2$ which forms H$_3^+$ via H$_2^+$+H$_2\rightarrow$H$_3^+$+H and ${k_{\rm H^+_3}(T)}$ is the reaction rate coefficient for the main destruction mechanism of H$_3^+$. We assume that H$_3^+$ is predominantly destroyed by the reaction in Eq.~\ref{eqn:HCO_form_react} \citep{Oka_H3_06}, i.e. we set $k_{\rm H^+_3}(T)$ equal to $k(T)$. 

%\citet{Oka_H3_06} \st{assumes a constant value for $k(T)$ of 2$\times$10$^{-9}$\,cm$^3$\,s$^{-1}$. We investigate the change in $k(T)$ across our chosen timescale by approximating how the remnant's temperature changes with time via: $T_{\rm gas}(M_{\rm C+O}, d_{\rm SN}) = T_{\rm gas}(M_{\rm C+O}, 100)\times(d_{\rm SN}/100)^{-1.26}$ }\citep{sarangi_cherchneff}. \st{We take $T_{\rm gas}(M_{\rm C+O}, 100)$, which is the temperature of the gas at the carbon + oxygen zone in the ejecta at $d_{\rm SN}$=100, taken to be 7580\,K } \citep[See Table 2, Zone 4B in ][]{sarangi_cherchneff}. \st{This leads to a change in temperature from 417\,K at $d_{\rm SN}$=1,000 to 20\,K at $d_{\rm SN}$=11,499, which changes $k(T)$ from 1.3$\times$10$^{-9}$\,cm$^3$\,s$^{-1}$ at $d_{\rm SN}$=1,000 to 2.4$\times$10$^{-9}$\,cm$^3$\,s$^{-1}$ at $d_{\rm SN}$=11,499, thus showing that the assumption of a nearly constant $k(T)$ across our 30 year timescale is sufficient to calculate an order of magnitude estimate for the mass of HCO$^+$ formed in the ejecta.}
Substituting the $n_{\rm H_3^+}$ relation into $R_{\rm HCO^+}$ leads to: $R_{\rm HCO^+}=\zeta \times n_{\rm H_2}$. We adopt a canonical ISM value for $\zeta$ of 3$\times$10$^{-17}$\,s$^{-1}$ \citep{Oka_H3_06}.
%to begin with because the ionisation rate of the \textcolor{blue}{inner layers of the core, where we find CO and HCO$^+$} in SN 1987A is unknown. We discuss if this ISM value is appropriate for SN 1987A later.

We evolve the parameters $n_{\rm H_2}$ and $R_{\rm HCO^+}$ with time due to their dependence on the expansion of the ejecta. We take our $n_{\rm H_2}$ values in Sect.~\ref{RADEX} and volume of HCO$^+$ in the ejecta using the observed extent of the HCO$^+$ emission from Sect.~\ref{section:LE_measurements} to be the respective values of $n_{\rm H_2}$ and the ejecta volume at $d_{\rm SN}$=11,499. We evolve them back to $d_{\rm SN}$=1,000 assuming a homologous $t^{-3}$ relationship, which accounts for the expansion velocity of the ejecta, leading to smaller ejecta volumes at earlier days.

%We note that our method to estimate the extent of the HCO$^+$ emission using the aperture described in Sect.~\ref{section:LE_measurements} will overestimate the volume of HCO$^+$ as, in reality, the emission is more likely to be toroidal/ellipsoidal shape such as that observed in CO (e.g. \citealt{Abellan, Cigan_ref}). To remedy this, we multiply the HCO$^+$ volume by the filling factors of CO found in \citet{Mikako_ref} assuming HCO$^+$ and CO are spatially similar.

The estimated HCO$^+$ mass formed is $\sim$10$^{-7}$\,M$_\odot$ by $d_{\rm SN}$=11,499, this is an order of magnitude below our HCO$^+$ observations from which we measure a mass of 3$-$9$\times$10$^{-6}$\,M$_\odot$. We suspect this is predominantly due to the assumption that the rate of ionisation of H$_2$ is $\zeta=3$$\times$10$^{-17}$\,s$^{-1}$. In reality, we expect this value may be higher due to the ejecta being subject to energy deposition from X-rays originating from the ring which would be the dominant source of ionisation (e.g. \citealt{Fransson_2013}). \st{Additionally, }
Whereas in the ISM, the ionisation rate of H$_2$ is due to background cosmic rays (as is assumed in \citealt{Oka_H3_06}).
Nevertheless, to match our HCO$^+$ masses at $\sim$10$^{-6}$\,M$_\odot$, an increase of the ionisation rate by only an order of magnitude (3$\times$10$^{-16}$\,s$^{-1}$) is required.

A further caveat to our ALMA analysis  is that we assume a uniform density of HCO$^+$. This is not the case as seen from Fig.~\ref{fig:HCO+_4_fig} where the spatial distribution of HCO$^+$ is clumpy. We compensate for this by introducing a filling factor $f$ on the HCO$^+$ area, which is assumed to be 1. 
This means that the HCO$^+$ mass formed may be overestimated.

Our order of magnitude estimate of mass (10$^{-7}$\,M$_\odot$) is 10 orders of magnitude more than that estimated in \citet{R&W} who only considered HCO$^+$ formation in a hydrogen-poor environment in the ejecta. Therefore we suspect that HCO$^+$ forms more efficiently in the ejecta if a greater amount of hydrogen is present. 
However, as highlighted in modelling the formation of molecules in population III supernovae by \citet{Che_Dwe_09}, even if 10$\%$ of hydrogen is inwardly mixed to the inner layers of the ejecta, it greatly impacts the chemistry of the ejecta, leading to more hybrid molecules such as OH and H$_2$O to form. Therefore, full chemical network where alternative HCO$^+$ formation molecules such as CH$^+$, OH$^+$, H$_2$O, HCO, CO$_2$ and CO$^+$ as well as CO must be considered for a more accurate estimation of how HCO$^+$ forms in SN 1987A. Despite this, our simple estimation has found that HCO$^+$ forms a reasonable amount of mass over 30 years, thus, it is a possibility for HCO$^+$ to form from CO in the ejecta, provided hydrogen is also co-located with CO.

\subsection{Mixing}
From the rough estimation of HCO$^+$ mass in the previous section, we argue that there must be some amount of hydrogen in the ejecta for HCO$^+$ to form.  This can occur if it is inwardly mixed from the hydrogen-rich envelope surrounding the core. HCO$^+$ cannot form in an unmixed ejecta, indicated by \citet{R&W}, since the carbon and oxygen and hydrogen gases are in separate regions left behind by the structure of the progenitor star (e.g. \citealt{SN87A_abundances_woosley, Woosley_6_week_1988}) which is retained in an unmixed scenario. Conversely, we assume that CO and H$^+_3$ is uniformly mixed for our calculation of HCO$^+$ mass from its formation rate. This assumption is unrealistic given the partial retention of the progenitor structure and the spatially separate distribution of molecular clumps \citep[e.g.][]{Abellan, Cigan_ref}. 
Therefore the question still remains in regards to the extent of mixing the ejecta has undergone and what different types of mixing it has possibly been subjected to. We discuss firstly, a type of mixing known to occur in SN 1987A and then speculate that two alternative types of mixing could also occur which can enhance the HCO$^+$ formation, and thus lead to a greater HCO$^+$ mass. 

The first form of mixing, called macroscopic mixing, which is the large-scale transport of metals outwards and hydrogen inwards through different nuclear burning zones is known to occur in SN 1987A. It is caused by Rayleigh-Taylor instabilities at the interfaces of nuclear burning zones \citep{hammer_ref, 3d_model_ref, Utrobin_2015, Utrobin_2019, Gabler_2021, 2026MNRAS.tmp..301W} and can mix hydrogen from the envelope down to where CO resides in the ejecta. 

We speculate that two other forms of mixing on smaller scales could enable the HCO$^+$ formation in SN 1987A. One form of mixing which could occur alongside macroscopic mixing around the first few days of the SN explosion is microscopic mixing \citep{2006A&A...453..661K}, which can mix the atomic hydrogen, carbon and oxygen at the atomic level thus improving the chances to form HCO$^+$ when this microscopically mixed gas cools and turns molecular. Microscopic mixing in SN 1987A is met with problems however, as it cannot explain early observations such as the CO mass \citep{Lepp_1990, Liu+Dalgarno} and clumps of CO and SiO are distinctly separate in the ejecta \citep{Abellan, Cigan_ref} which argues against microscopic mixing. 

The final type of mixing which may potentially enhance the HCO$^+$ formation within the remnant is mixing that occurs in the progenitor star before going supernova (e.g. \citealt{Groh_presn_2019, Farrell_2021, Boccioli.2026}). Recent studies \citep{Groh_presn_2019, Farrell_2021} have found interactions between the He/H zones in BSG stars that could mix hydrogen into the helium zone of the star where small abundances of carbon and oxygen exist (e.g. \citealt{SN87A_abundances_woosley}) which can lead on to enhance the formation of HCO$^+$ after the star has gone supernova. A caveat to this is that only stars of very low metallicity (Z=0.0004\,Z$_\odot$) are considered for these studies. SN 1987A's progenitor is expected to have $Z\sim\dfrac{1}{2}$\,Z$_\odot$ (e.g. \citealt{SN87A_abundances_woosley, LMC_metallicity}), typical of the LMC, and the higher metallicity may affect how well this pre-SN mixing may occur. 
%\textcolor{magenta}{Furthermore, the modelling of turbulent convective mixing in pre-SN stars with main-sequence masses greater than 20\,M$_\odot$ have found that some of the helium can get mixed further in to the progenitor star where it gets burned into oxygen \citep{Frey_presn_mixing_2013}. We speculate that this can also enhance HCO$^+$ formation in the post-SN phase of the remnant, as this depletion of the helium layer between the hydrogen and C+O nuclear burning zones makes it easier for hydrogen to interact with C+O.}

\subsection{The Spatial Distributions of HCO$^+$, H$_2$ and H$\alpha$}
%We argue that there is still a dependence on $n_{\rm CO}$ in the HCO$^+$ formation reaction rate, $R_{\rm HCO^+}$, despite being no longer present in the equation. This is due to $n_{\rm H_2}$ being a tracer of $n_{\rm CO}$ in most regions of the ISM (e.g \citealt{Oka_H3_06}).
We argue that in order to form HCO$^+$, some H$_2$ must be co-located with CO. This is not \textbf{quite} reflected in recent observations of H$_2$ where its spatial distribution and \textbf{morphology} are different to that of CO and HCO$^+$ \citep[Fig.~\ref{fig:H_alpha_HCO}: ][]{2019_Larsson_H2, Matsuura_2024_H2}. Instead, the H$_2$ emission more closely resembles the H$\alpha$ morphology. We speculate however, that some H$_2$ may still be co-located with CO but would not be detected in the near-infrared images of \citet{2019_Larsson_H2} and \citet{Matsuura_2024_H2}, because the distribution of the H$_2$ emission in these observations 
reflect the location of the mechanisms which power its emission rather than the presence of H$_2$ itself. 
%What causes the excitation of H$_2$ in the ejecta was investigated by \citet{2016_fransson_H2_discovery}, where H$_2$ could be excited from either UV emission arising from the ring or from collisional excitations by electrons. Since neither are used to excite the HCO$^+$ nor CO emission, then there are few similarities between the distributions of the H$_2$ emission with that of CO and HCO$^+$ emission.

In Sect.~\ref{disc:chemistry}, we considered only the formation of HCO$^+$, discussing whether this formation process could potentially produce an amount of HCO$^+$ comparable to the observed value over the lifetime of the system. In reality, however, HCO$^+$ is also subject to destruction processes. We therefore determine whether and where HCO$^+$ can survive at the present day by considering the balance between its formation and destruction rates. This chemical balance determines the HCO$^+$-emitting region.
The dominant destruction pathway for molecular ions is typically dissociative recombination with free electrons. For HCO$^+$, the UMIST rate coefficient \citep{UMIST_2024} is $\sim 2 \times 10^{-7}\,\mathrm{cm^3\,s^{-1}}$, which is about two orders of magnitude faster than the HCO$^+$ formation rate considered in Sect.~\ref{disc:chemistry} ($\sim 2 \times 10^{-9}\,\mathrm{cm^3\,s^{-1}}$).
This imbalance likely explains the observed morphological differences between species. In particular, HCO$^+$ can only form efficiently when the CO + H$_3^+$ reaction outpaces the destruction of both HCO$^+$ and H$_3^+$ by free electrons. This condition requires a low ionisation fraction, or low electron density, very roughly, $\sim$100 times lower than the CO abundance. Consequently, HCO$^+$ is not expected to coexist with H$\alpha$ emission, which traces ionised hydrogen.

The lack of co-spatiality between H$_2$ and HCO$^+$ may provide insight into the excitation mechanism of H$_2$ in the ejecta.  \citet{2016_fransson_H2_discovery} proposed that H$_2$ excitation could arise either from UV irradiation originating in the circumstellar ring or from collisional excitation by electrons. The latter scenario would enhance the destruction of H$_3^+$ and HCO$^+$, hence H$_2$ and HCO$^+$ would not be co-spatial. %while the former could contribute to CO dissociation and increase the free electron population via photoionisation of carbon.
Hence, UV irradiation is favoured.  This interpretation is consistent with the more recent H$_2$ excitation model of \citet{Larsson_model_2023}.

%%%%%%%%%%%%%%%%%%%%%%%%%%%%%%%%%%%%%%%%%%%%%%%%%%%%%%%%%%%%%%%%%

\section{Conclusions}
We analysed the line emissions of the $J=3-2$ and $4-3$ transitions of HCO$^+$ at
265.6$-$269.2 and 354.7$-$358.0\,GHz, respectively, using the ALMA to investigate how it can form in the ejecta of SN 1987A and to speculate on the extent of chemical mixing the ejecta has undergone to enable HCO$^+$ formation.
The spatial distributions of $J=3-2$ HCO$^+$ line emission was compared to that of the $J=2-1$ CO line emission which were found to have similar distributions. The spatial distributions of HCO$^+$ and CO  were found to be correlated, with a strength of 0.72. The degree of ionisation that the HCO$^+$ gas is subjected to can also be investigated by comparing the $J=3-2$ HCO$^+$ spatial distribution to that of H$\alpha$. It was found that HCO$^+$ is more compact in the ejecta compared to H$\alpha$ thus indicating that HCO$^+$ forms deeper within the ejecta where ionisation levels are milder.
The mass of HCO$^+$ is calculated using HCO$^+$ line intensities with the radiative transfer code {\sc RADEX}. This returned a HCO$^+$ mass of (2.94 $\,^{+5.7}_{-1.2}$)$\times$10$^{-6}\,$M$_{\odot}$ and (8.80$\,^{+31.5}_{-4.7}$)$\times$10$^{-6}\,$M$_{\odot}$ depending on the choice of H$_2$ collisional partner density; 10$^6$ or 10$^5$\,cm$^{-3}$ respectively. We additionally calculate an order of magnitude estimate for the mass of HCO$^+$ formed from a single HCO$^+$ formation reaction involving CO; namely CO+H$_3^+\rightarrow$\,HCO$^+$+H$_2$ across $\sim$30 years to investigate if its formation from CO is realistic. This calculation forms $\sim$10$^{-7}$\,M$_\odot$ of HCO$^+$ which is an order of magnitude below the HCO$^+$ mass based on current observations. We suspect that the underestimation of mass stems from the uncertainty in the ionisation rate of H$_2$ which we also suspect to be underestimated.
We also suggest that, in order to replicate the observed HCO$^+$ masses of $\sim$10$^{-6}$\,M$_\odot$, a moderate amount of H$_2$ needs to co-located with the CO. We speculate, that as well as macroscopic mixing, which caused the hydrogen to inwardly mix to the carbon-rich layers of the core where we find CO, additional forms of mixing on smaller scales could occur to enhance the HCO$^+$ formation. This may be of the form of microscopic mixing at the molecular level between CO and H$_2$ and/or mixing which has taken place between the He/H layer in the progenitor star before it exploded as SN 1987A. This study's findings need to be confirmed by more extensive chemical networks for supernova remnants which involve more hydrogen chemistry. Our analysis of HCO$^+$ however, has shown that the ejecta of SN 1987A still offers new avenues of chemistry and hydrodynamics to explore.

\section*{Acknowledgements}
HD acknowledge support from STFC for her PhD studentship (2422911). MM acknowledges support from the STFC Ernest Rutherford fellowship (ST/L003597/1) and STFC Consolidated grant (ST/W000830/1). HLG acknowledges support from the European Research Council (ERC) in the form of Consolidator Grant {\sc CosmicDust} (ERC-2014-CoG-647939).
%MMe acknowledges that  a portion of their research was carried out at the Jet Propulsion Laboratory, California Institute of Technology, under a contract with the National Aeronautics and Space Administration (80NM0018D0004).
This paper makes use of the following ALMA data:
ADS/JAO.ALMA\#2016.1.00077.S, ADS/JAO.ALMA\#2017.1.00221.S,  ADS/JAO.ALMA\#2013.1.00063.S, ADS/JAO.ALMA\#2013.1.00280.S and
ADS/JAO.ALMA\#2021.1.00707.S.
ALMA is a partnership of ESO (representing its member states), NSF (USA) and NINS (Japan), together with NRC (Canada), MOST and ASIAA (Taiwan), and KASI (Republic of Korea), in cooperation with the Republic of Chile. The Joint ALMA Observatory is operated by ESO, AUI/NRAO and NAOJ. 
Support for HST GO program numbers 13401, 13405 and 15256 was provided by NASA through grants from the Space Telescope Science Institute, which is operated by the Association of Universities for Research in Astronomy, Inc., under NASA contract NAS5-26555.
This work is based on observations made with the NASA/ESA/CSA James Webb Space Telescope. The data were obtained from the Mikulski Archive for Space Telescopes at the Space Telescope Science Institute, which is operated by the Association of Universities for Research in Astronomy, Inc., under NASA contract NAS 5-03127 for JWST. These observations are associated with program \#1726.
This research has made use of {\tt Astropy\footnote{\url{http://www.astropy.org}}}, a community-developed core Python package for Astronomy (\citealt{Astropy2013,Astropy2018, Astropy2022}), {\tt CASA}, a software pipeline designed for data processing of ALMA and VLT data (\citealt{The_CASA_Team_2022}) and the Python libraries {\tt SciPy} (\citealt{Scipy2020}), 
{\tt NumPy} (\citealt{Numpy}) and {\tt Matplotlib} (\citealt{Matplotlib})
\\

%%%%%%%%%%%%%%%%%%%%%%%%%%%%%%%%%%%%%%%%%%%%%%%%%%
\section*{Data Availability}
The ALMA data products are available on the ALMA archive \url{https://almascience.eso.org/alma-data} or via request through the corresponding author.


\bibliographystyle{mnras}
\bibliography{example} % if your bibtex file is called example.bib

%\begin{thebibliography}{99}
%\end{thebibliography}

%%%%%%%%%%%%%%%%%%%%%%%%%%%%%%%%%%%%%%%%%%%%%%%%%%

%%%%%%%%%%%%%%%%% APPENDICES %%%%%%%%%%%%%%%%%%%%%

%\begin{appendices}
\appendix
\section{Appendix}
\label{app:sio}

%HCO$^+$ with SiO
\begin{figure*}
	\includegraphics[clip, trim=2.3cm 0.5cm -0.8cm 1.9cm, width=24cm]{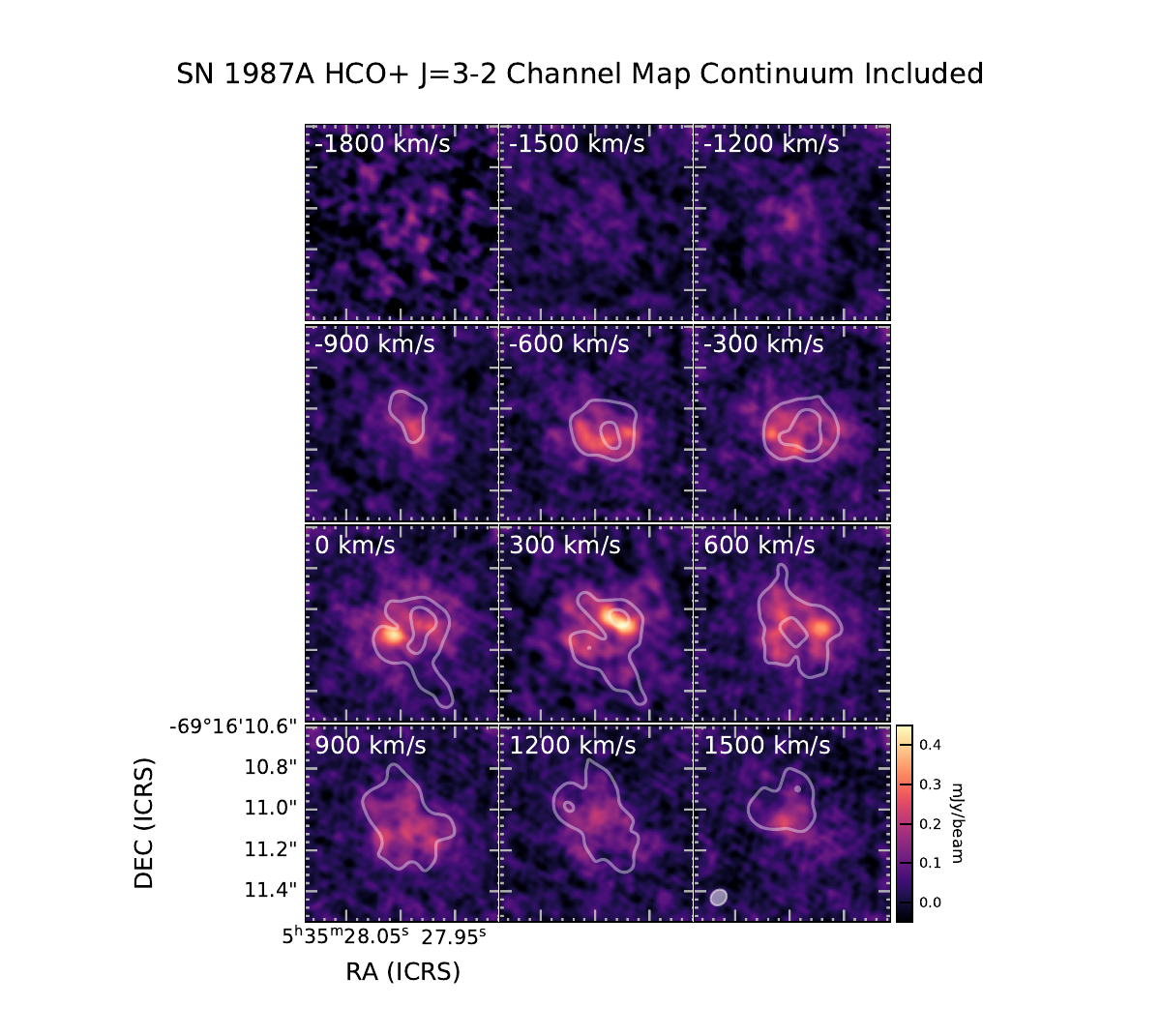}
    \caption{Similar to Figure~\ref{fig:HCO+_4_fig}, here we show the HCO$^+$ $J=3-2$ channel map with the continuum-included in the brightness distributions. Line contours show the spatial distribution of SiO $J=5-4$. The contour levels at levels of 0.5 and 1.2\,mJy\,beam$^{-1}$. These levels were chosen to be consistent with the CO levels in Figure~\ref{fig:HCO+_4_fig} and also to best represent the SiO spatial distribution. The SiO distribution is $\sim$\,4.4 times brighter than the HCO$^+$ distribution if beam size and pixel scale are the same in the HCO$^+$ and SiO channel maps. The overall spatial distributions of the two species differ greatly, but there are overlapping areas of higher emission corresponding to the channels 0\,km\,s$^{-1}$ and 300\,km\,s$^{-1}$. All enclosed SiO contours are bright spots except for the large ellipse in the centre of the SiO distribution corresponding to the $-$600\,km\,s$^{-1}$ channel which is in fact, a hole. See \citet{Cigan_ref} for more details. }
    \label{appendix:SiO_figure}
\end{figure*}

%J=4-3 channel map 
\begin{figure*}
    \includegraphics[clip, trim=2.5cm 0.7cm 0.2cm 1.5cm, width=22cm]{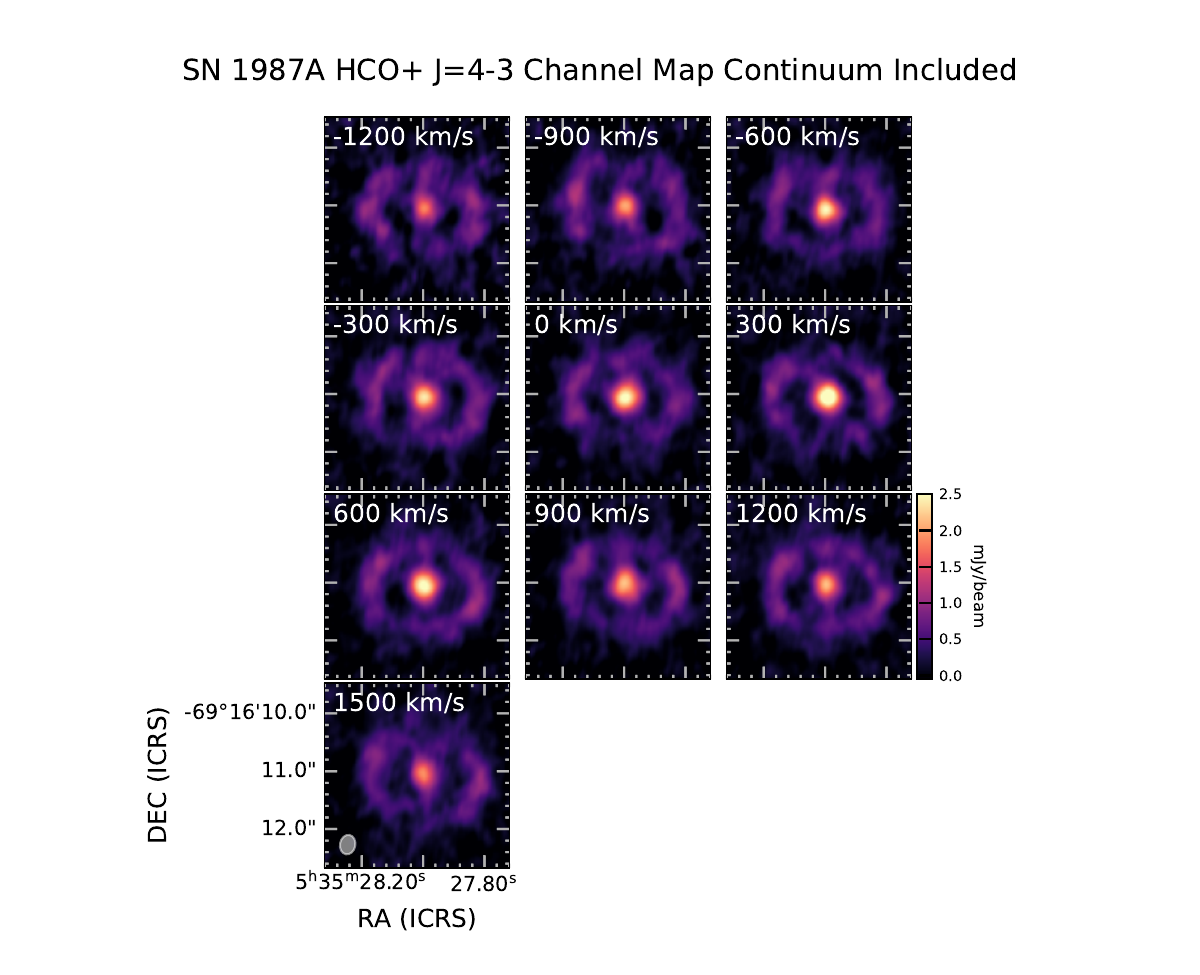}
    \caption{The velocity channel map of the continuum-included $J=4-3$ transition of HCO$^+$ from velocities of $-$1350\,km\,s$^{-1}$ to 1650\,km\,s$^{-1}$, binned in 300\,km\,s$^{-1}$ increments. Each image channel is labelled with the central velocity of the 300\,km\,s$^{-1}$ bin. Velocities are in LSRK and are respective to the rest frequency of the $J=4-3$ HCO$^+$ transition. The ejecta emission from this region is not spatially resolved as seen with the brighter `blob' in the centre of the ring. Synchrotron emission from the ring surrounding the ejecta is also present in this channel map. %\textcolor{magenta}{Velocities are in LSRK and have not been corrected for the systemic velocity of SN 1987A at 286.7\,km\,s$^{-1}$ \citep{Sys_vel_1987A_2008}}.
    }
    \label{appendix:HCO_43}
\end{figure*}

\newcommand*{\MAGMA}{\textsc{magma}\xspace}
%%%%%%%%%%%%%%%%%%%%%%%%%%%%%%%%%%%%%%%%%%%%%%%%%%

% Don't change these lines
\bsp	% typesetting comment
\label{lastpage}
\end{document}